\documentclass[%
 reprint,
 amsmath,amssymb,
 aps,
floatfix
]{revtex4-2}
\usepackage[utf8]{inputenc}
\usepackage{graphicx}
\usepackage{dcolumn}
\usepackage{bm}
\usepackage{array}
\usepackage{hyperref}
\usepackage{amsmath}
\usepackage[mathlines]{lineno}
\usepackage{amssymb}
\usepackage{textcomp,gensymb}
\usepackage[greek,main=english,german]{babel}
\usepackage{float}
\begin{document}

\preprint{APS/123-QED}

\title{Direct-write laser-induced self-organization and metallization beyond the focal volume in tellurite glass}

\author{G{\"o}zden Torun\textsuperscript{1}}%
 \email{gozden.torun@epfl.ch}
\author{Tetsuo Kishi\textsuperscript{2}}
\author{Yves Bellouard\textsuperscript{1}}
\affiliation{\textsuperscript{1}Galatea Lab, STI/IMT, Ecole Polytechnique Fédérale de Lausanne (EPFL), 2002 Neuchâtel, Switzerland}%
\affiliation{\textsuperscript{2}Department of Chemistry and Materials Science, Tokyo Institute of Technology, 2-12-1 Ookayama Meguro-ku, Tokyo, Japan}%

\date{\today}

\begin{abstract}
We report on a self-organized process initiated on a potassium-tungsten-tellurite glass surface under repetitive femtosecond laser irradiation in a regime where cumulative effects lead to a localized melting. Specifically, we show that self-organized periodic patterns consisting of parallel nanoplanes perpendicular to the laser polarization are forming and extending beyond the zone under direct laser exposure. Examination of the modified regions revealed a phase change from a glassy tellurium-oxide to a crystalline elemental tellurium. In addition, we observe that this self-organization process, associated with elemental redistribution and deoxygenation, is triggered by the optical-field strength. We suggest that early self-organized nanostructures formed by a local-field enhancement is subsequently reinforced by a metallization event in an open-air atmosphere.
\end{abstract}

\maketitle


\section{\label{sec:level1}INTRODUCTION}

Over the past decades, increasing research activities investigating the formation of periodic patterns on various surfaces exposed to intense pico- and femtosecond laser beams were conducted. This research effort has recently been summarized by J. Bonse \textit{et al.} \cite{bonse_femtosecond_2012}. Grating-like patterns have been observed at the surfaces of various types of materials, such as metals, semiconductors, and dielectrics \cite{bonse_femtosecond_2012}, suggesting the existence of generic mechanisms causing their formations. Due to their spatial micron to sub-micron periodicities, these nanostructures offer efficient means for controlling surface properties. For instance, one can use surface periodic patterns for tailoring wetting properties from hydrophilic to hydrophobic behavior, for encoding complex color interference-based patterns, or for modifying wear-rate in tribological applications \cite{vorobyev_colorizing_2008, baldacchini_superhydrophobic_2006, eichstadt}. However, the formation mechanisms of these periodic surface nanostructures remains controversial \cite{bonse2020}. Interference effects along with transient changes in the optical properties of the surface \cite{wu_femtosecond_2003}, second harmonic generation (SHG) \cite{borowiec_subwavelength_2003, jia_formation_2005, bonse_structure_2005}, excitation of surface plasmon polaritons (SPPs) \cite{martsinovskii_ultrashort_2008, miyaji_origin_2008} or self-organization \cite{shugaev_mechanism_2017,reif_ripples_2002} are just a few examples of possible mechanisms that have been debated to explain these phenomena. In addition, chemical alterations such as a surface oxidation due to the irradiation in an air environment \cite{li_formation_2014} and a phase transformation \cite{costache_subdamagethreshold_2004} were observed in the regime of periodic surface structures. In attempts to identify generic principles underlying the formation of these nanostructures, previous studies explored the role of specific material properties such as viscosity \cite{graf_formation_2017}, optical refractive index \cite{ding_micro-raman_2009}, band-gap \cite{borowiec_subwavelength_2003}, and structural defects \cite{ly_role_2017}.

In this work, we investigate the formation of self-organized nanostructures on the surface of the tellurite glass resulting from a multi-pulse irradiation coming from a scanning femtosecond laser beam. In particular, we report on new observations that challenges our current understanding of laser-induced nanogratings formation, and that highlight the complexity and diversity of phenomena underlying this self-organization process. The choice of the tellurite glass as a test material is motivated by its high technological relevance in a broad number of fields, and its metastable glass behavior that makes it easily prone to crystallization. Tellurium-based systems are subjected to a wide range of reversible and irreversible phase changes \cite{wuttig_phase-change_2007, matsunaga_local_2011, li_pressure-induced_2010, celikbilek_crystallization_2011}. Especially, high refractive index and third-order nonlinear susceptibility, high dielectric constant, showing SHG, low glass transition and phonon energies are essential for technological applications such as all‐optical switch, photonic diode, re-writable high-density storage media and optical recording \cite{wuttig_phase-change_2007, matsunaga_local_2011, li_pressure-induced_2010, celikbilek_crystallization_2011, el2011tellurite, kosuge_thermal_1998, celikbilek_ersundu_evaluation_2017, sidkey_ultrasonic_2004, tanabe_upconversion_nodate, wang1194, nasu1990,wu2019}. Among various TeO\textsubscript{2}-based glasses, K\textsubscript{2}O - WO\textsubscript{3} - TeO\textsubscript{2} has a wide glass-forming region, and its properties have been studied well \cite{el2011tellurite, kosuge_thermal_1998, celikbilek_ersundu_evaluation_2017, sidkey_ultrasonic_2004}. In this particular case, we demonstrate a self-organization phenomenon driven by local-field enhancement and controlled by the electric field that spans beyond the focal volume, and that leads to the formation of thin metallic surface accompanied by self-organized nanoplanes embedded in a dielectric matrix.

\section{\label{sec:level2}EXPERIMENTAL PROCEDURES}

\subsection{Glass specimen preparation}
The nominal composition of the glass used in this study is 10K\textsubscript{2}O-10WO\textsubscript{3}-80TeO\textsubscript{2} (mol\%). Commercial powders of reagent K\textsubscript{2}CO\textsubscript{3} (99.5\%), WO\textsubscript{3} (99\%) and TeO\textsubscript{2} (99\%) were mixed and melted in Au crucible at around 973K for 30 minutes in an electric furnace. The melt was quenched onto a brass plate. After quenching, the specimen was crushed and remelted at 973K for 30 minutes, which was followed by a subsequent annealing at 598K for 1 hour. The specimen was cut and polished prior to femtosecond laser exposure. 
\subsection{Femtosecond laser inscription}
An Yb fiber-amplifier femtosecond laser (Yuzu from Amplitude Laser) emitting 270 fs pulses at 1030 nm was used in this experiment. Laser patterns were inscribed on the surface of potassium-tungsten tellurite glass as a line with a length of 10 mm. The specimen was translated under the laser focus with the help of a high precision motorized stage (Ultra-HR from PI Micos). The laser beam was focused on the surface of the specimen with a 0.4 numerical aperture (NA) objective (OFR-20x-1064 nm from Thorlabs), resulting in a spot-size (defined at 1/e\textsuperscript{2}) of 1.94 $\mu$m (see Supplementary Material for the laser spot size measurement). The transition from non-cumulative to a cumulative thermal regime exposure regime was determined by observing the evolution of the width of the laser affected zone with the pulse repetition frequencies for a same net fluence and was found around 300-400 kHz. In what follows, the repetition rate was fixed at 1 MHz, which lies well within the cumulative regime for this particular glass. For comparative experiments between the two exposure regimes, 100 kHz-pulse repetition rate was used for representing the non-cumulative thermal exposure regime. Pulse energy and translation velocity were selected as main variables to obtain different deposited energies. Accordingly, the deposited energy per unit surface (or ‘net fluence’) on the specimen can be approximated by
\begin{equation}
E\textsubscript{deposited}=\frac{4E\textsubscript{p}}{\pi \omega} (\frac{f}{v})
\end{equation}
where E\textsubscript{p} is the pulse energy, ${\omega}$ is the optical beam waist (defined at 1/e\textsuperscript{2}), f is the laser repetition rate and v is the writing speed as described in Ref. \cite{Rajesh}. This deposited energy is a dose of how much energy per unit surface is passing through the material but does not indicate how much of this energy is effectively absorbed. However, it remains a convenient and simple metric for comparing and reproducing experimental exposure conditions. Here, pulse-to-pulse overlapping ratios were varying from 95 to 99.9\%, and the pulse energy was ranging from 1 nJ to 200 nJ.
Furthermore, tracks were inscribed using opposite directions of laser beam movement along a single writing axis and under three different linear polarization states (and therefore, orientation of the electrical field E) defined as parallel, at forty five degree, and perpendicular to the writing direction, respectively. 
\subsection{Specimen characterization}
After laser exposure, specimens were first observed using an optical microscope (BX51 from Olympus), and subsequently coated with carbon thin film for high-resolution imaging and elemental analysis using a field-emission scanning electron microscope (FE-SEM, Gemini 2 from Zeiss) equipped with energy-dispersive X-ray spectroscopy (EDS) operated at 20 kV. A cross-sectional milling and observations were performed in another scanning electron microscope combined with focused ion beam (SEM-FIB from FEI Nova 600 NanoLab). The milling was performed using a Ga-ion source operated at 30 kV acceleration voltage and 7 nA emission current. A Raman spectrometer (LabRam HR from Horiba) equipped with a 532 nm-laser excitation source attenuated down to 4 mW (to prevent damaging the specimen) was used to record Raman spectra of the modified regions. The linearly-polarized Raman laser beam was focused at the surface of the specimen using a 0.9 numerical aperture (NA) objective (100x-532 nm from Thorlabs). A series of line scans were performed at room temperature on each laser-modified regions, from 10 $\mu$m outside the laser-written lines towards the center of the modification with a period of 1 $\mu$m, and with acquisition times of 30 seconds for each individual spots.

\section{\label{sec:level3}RESULTS AND DISCUSSION}

\subsection{Self-organized nanostructures beyond the focal volume}

Figure 1 shows the laser-affected zones after irradiation with a pulse energy of 200 nJ and a pulse overlapping ratio of 99.5\% at a pulse repetition rate of 1 MHz (a-g) and 100 kHz (h). At 1 MHz (Fig. 1, a-g), in the thermal cumulative regime, self-organized nanostructures, consisting of parallel planes oriented perpendicular to the laser electrical field orientation, significantly wider than the focal area of the laser beam formed at the tellurite glass surface are observed. This fact raises interesting questions concerning the underlying mechanism driving self-organization in this particular case. To date, the formation of nanogratings has been interpreted as a subtle interplay between the incoming intense laser field and the material under direct exposure. While such a mechanism may also be present here, it is not sufficient in itself to explain the full extent of the modification.

The modification exhibits two distinctive periodic arrangements: one, in the center, characterized by 800 nm-spaced nanoplanes, and a second one, near the edges of the laser modified zone, consisting of 250 nm-nanogratings with a much smaller span than the one found in the center. It is worth noticing that the secondary nanostructures found at the edges of laser-affected zones are best visible for a polarization perpendicular to the writing direction, and absent for a polarization aligned with the writing axis.

\begin{figure*}
\centering
\includegraphics[width=18cm]{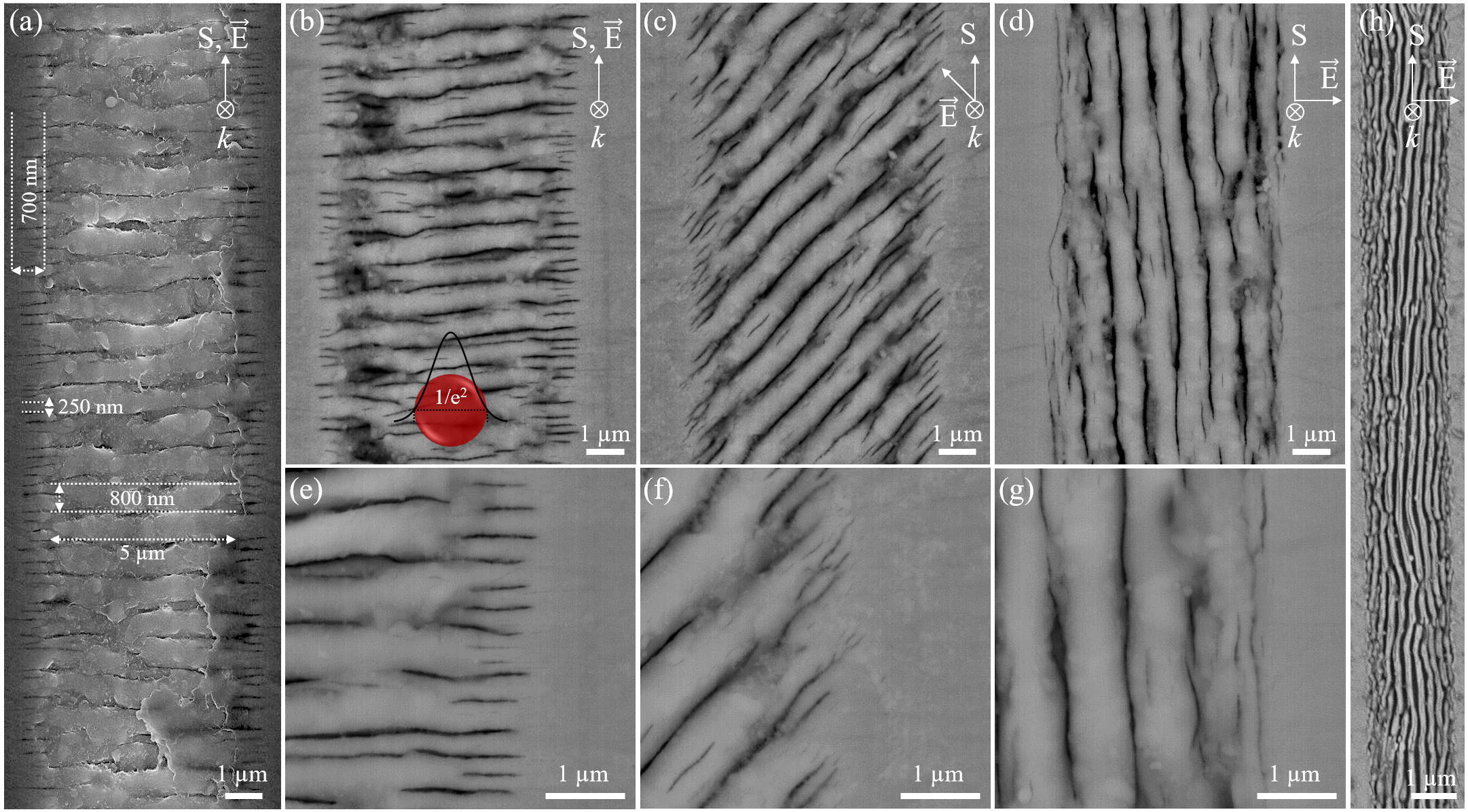}
\caption{\label{fig:wide}Secondary electron (SE) and backscattered electron (BSE) images of potassium-tungsten tellurite surface after femtosecond laser irradiation at 1 MHz with different linear polarization orientations with respect to writing direction (as indicated with the electric field orientation): (a) Secondary electron (SE) and (b) backscattered electron (BSE) images with longitudinal polarization, (c) 45-degree polarization and (d) transverse polarization. (e-g) Magnified BSE images of a boundary between the modified region and the parent material. The incoming pulse energy and writing speed were 200 nJ and 10 mm/s (incoming pulse fluence: 13 J/mm\textsuperscript{2}), respectively. The red circle indicates the diameter of the laser spot. The laser modified zone containing nanostructures stretches significantly outside the zone that was under direct laser exposure. (h) SE image of potassium-tungsten tellurite surface after femtosecond laser irradiation at 100 kHz (200 nJ, 10 mm/s) for comparison. The width is measured as 1.8 $\mu$m in this non-cumulative regime, which is in agreement with the measured optical beam diameter. The laser was propagating perpendicular to the image plane. The writing direction was from the bottom to the top of the image.}
\end{figure*}

The electric field strength appears to be an essential parameter for triggering or suppressing self-organized nanostructures at the focal area as evidenced by varying the pulse energies in Figure 2. When multiple pulses accumulate at a fast enough rate, the temperature rise causes the material under exposure to melt locally (Fig. 2(a)). The nanoplanes and the secondary nanostructures are forming at higher pulse energies, and consequently at higher temperatures. They are not found at lower pulse energies (Fig. 2(b)). This sharp transition suggests the existence of a field-intensity threshold for triggering the self-organization process. In another research field exploring ‘electro-crystallization’, Luedtke \textit{et al.} showed field-induced shape deformations in dielectric liquid nano-droplets \cite{luedtke_dielectric_2011}. Similar to Figure 2(c), at low electric field, field-induced shape change was observed. Further increase in the electric field resulted in a gradual enhancement of the molecular dipole reorientation. Although in Luedtke \textit{et al.}, the electrical field is not produced by a laser, we note intriguing similarities.

\begin{figure}
\centering
\includegraphics[width=8.5cm]{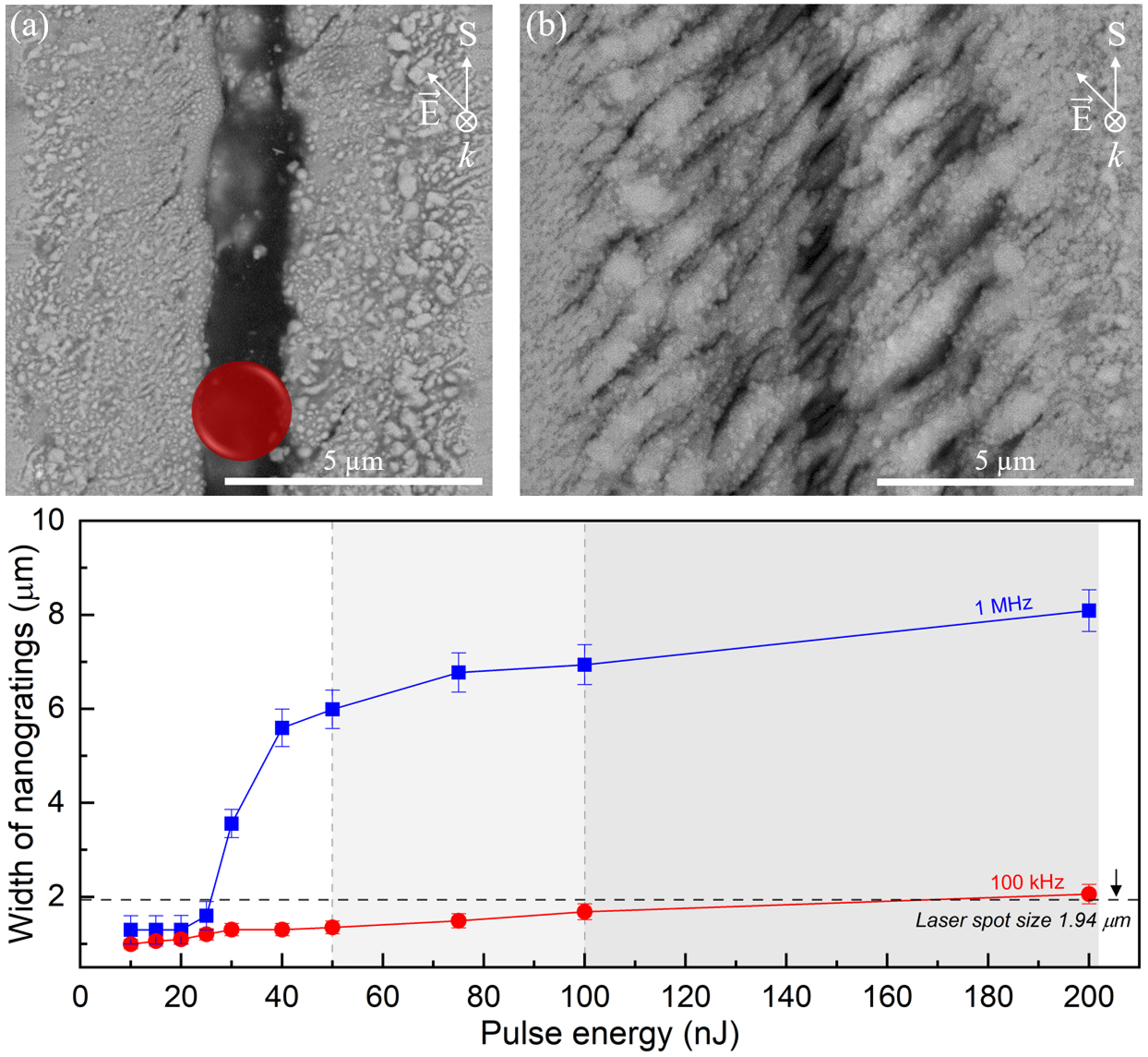}
\caption{\label{fig:epsart}Top: BSE images of potassium-tungsten tellurite glass surface after femtosecond laser irradiation with two different irradiation parameters: (a) 10 nJ, 0.5 mm/s (13 J/mm\textsuperscript{2}); (b) 200 nJ, 0.5 mm/s (262 J/mm\textsuperscript{2}), both obtained in the thermal cumulative regime. Bottom: Plot of the width of the microstructures versus pulse energy at constant writing speed (0.5 mm/s). In the cumulative regime (blue curve at 1 MHz), a transition between two different modifications is observed. Below 30 nJ, an homogeneous molten region is found within the laser waist region. Above this value, self-organized structures consisting of parallel nanoplanes spaced by about 800 nm are present and gradually expand well beyond the zone under laser exposure. Both images correspond to 45$^{\circ}$ polarization (with respect to the writing direction). For comparison, the curve in red shows the width of the modified microstructures in the case of the non-cumulative regime (red curve at 100 kHz). In the non-thermal cumulative process, the width remains within the range defined by the beam waist at the surface.}
\end{figure}

In practice, the observable surface modification-threshold is found for pulse energies of 10 nJ, while the nanostructures formation starts at 20 nJ and extends beyond 200 nJ, up to the point where ablation occurs. As a characteristic signature of a temperature-cumulative process, decreasing the translation velocity (i.e. increasing the pulse-to-pulse overlapping ratio) at fixed pulse energy leads to an increase of the modified area width. Likewise, at fixed translation velocity, increasing pulse energy causes a widening of the modified area up to 10 $\mu$m. The span of self-organized nanostructures change accordingly with an exposure time and space, and remarkably, beyond the region under direct laser-exposure. 

FIB milling was performed to observe the profile of these nanostructures and their extent within the material itself (Figure 3). Unlike surface structures reported before that are in the scale of a few tens of nanometers, the depth of the nanostructures is in our case around 1 $\mu$m and increases with increasing the pulse energy \cite{bonse_laser-induced_2017}. These observations show that higher fluence increase \textit{both} lateral and depth extensions of modified volumes. A fluence-dependent growth of the melt depth has been observed in photo-excited tellurium films \cite{pamler_transient_1987, cheng_femtosecond_2018}.
\begin{figure}
\centering
\includegraphics[width=8.5cm]{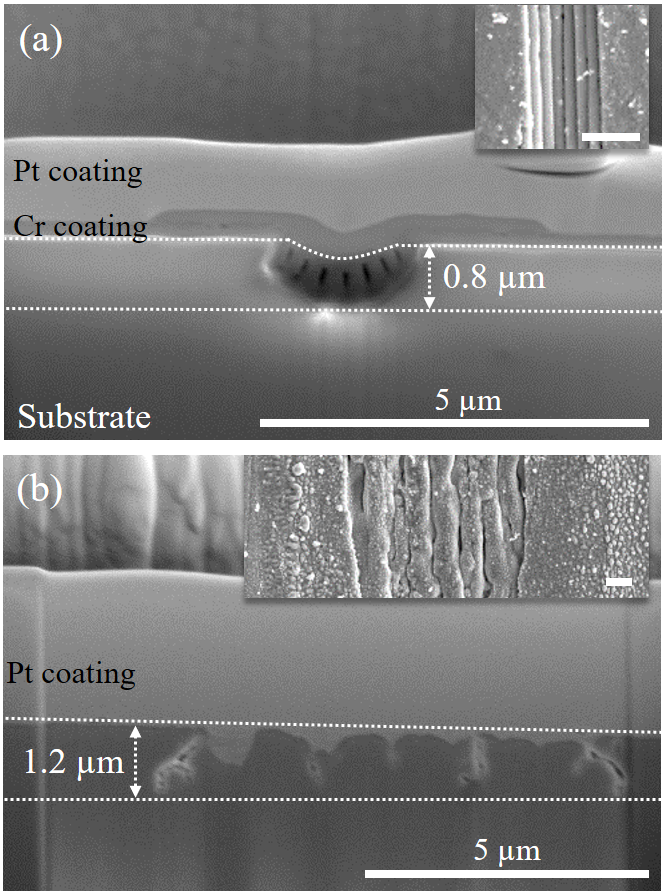}
\caption{\label{fig:epsart} SE images of FIB-cut in potassium-tungsten tellurite glass: (a) 30 nJ, 100 mm/s (0.2 J/mm\textsuperscript{2}) and (b) 100 nJ, 0.5 mm/s (131 J/mm\textsuperscript{2} at 1 MHz), respectively. Both images correspond to a transverse polarization (with respect to the writing direction) and 1 MHz pulse repetition rate. Inset images show the modified zone before FIB process. The scale bars in both images are 1 $\mu$m.}
\end{figure}
\subsection{Elemental analysis and bond characteristics}
\begin{figure}
\centering
\includegraphics[width=8cm]{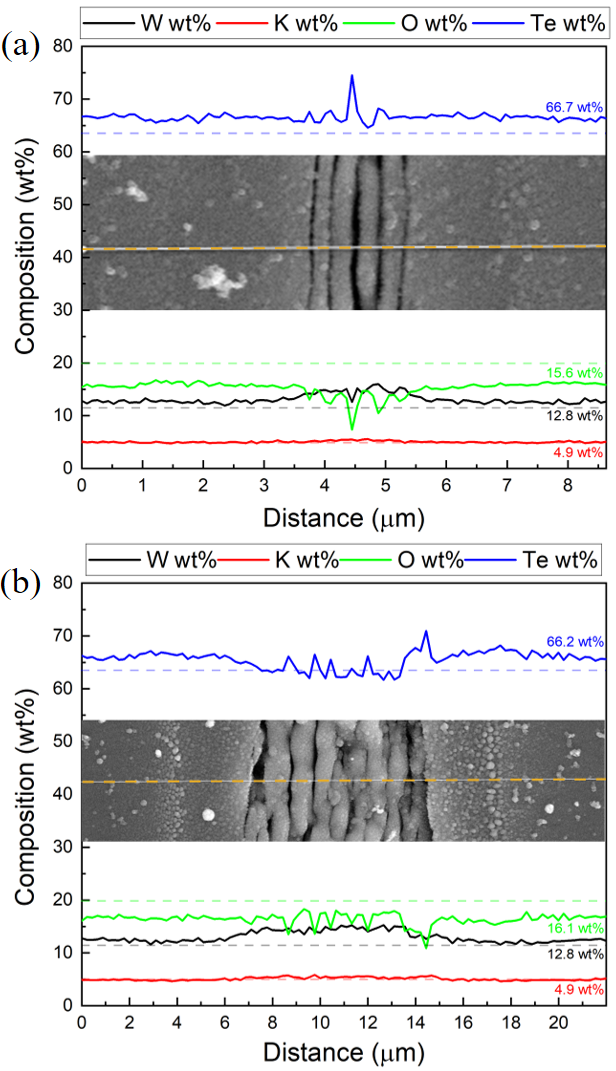}
\caption{\label{fig:epsart}Elemental distribution across a modified zone accompanied with SE images at 1 MHz: (a) 20 nJ, 0.5 mm/s (26 J/mm\textsuperscript{2}); (b) 100 nJ, 0.5 mm/s (131 J/mm\textsuperscript{2}). Colored values indicate the average composition along the scanned line. Dashed lines indicate the composition of pristine glass.}
\end{figure}
Energy dispersive spectroscopy (EDS) observations of laser modified zones in potassium-tungsten tellurite glass are shown in Figure 4. Overall, small variations in the chemical composition have been observed as a consequence of laser modifications. The first nanostructures observed for a laser-pulse energy of 20 nJ is shown in Figure 4(a). Nanoplanes show alternating tellurium (Te) and oxygen (O) contents, whereas the overall tungsten (W) content remains slightly higher. Figure 4(b) shows elemental distribution across a laser modified zone obtained with a pulse energy of 100 nJ. Similarly, alternating Te and O contents along nanoplanes were observed. Compared to the pristine material (dashed lines in Fig. 4), both modifications show approximately 3 wt\% higher Te, 4 wt\% lower O and 1 wt\% higher W contents. Changes in chemical composition for tungsten, tellurium and oxygen elements are also found beyond the nanoplanes. On the other hand, the content of potassium (K) atoms remains constant since it is the key modifier of the backbone tellurite glass structure. Experiments have shown that potassium is not sensitive to variations in local structures and as such, the structure of potassium-tellurite (K\textsubscript{2}O-TeO\textsubscript{2}) glass has been preserved without any decomposition during heating and cooling cycles \cite{Akagi_1999}.

\begin{figure}
\centering
\includegraphics[width=8.5cm]{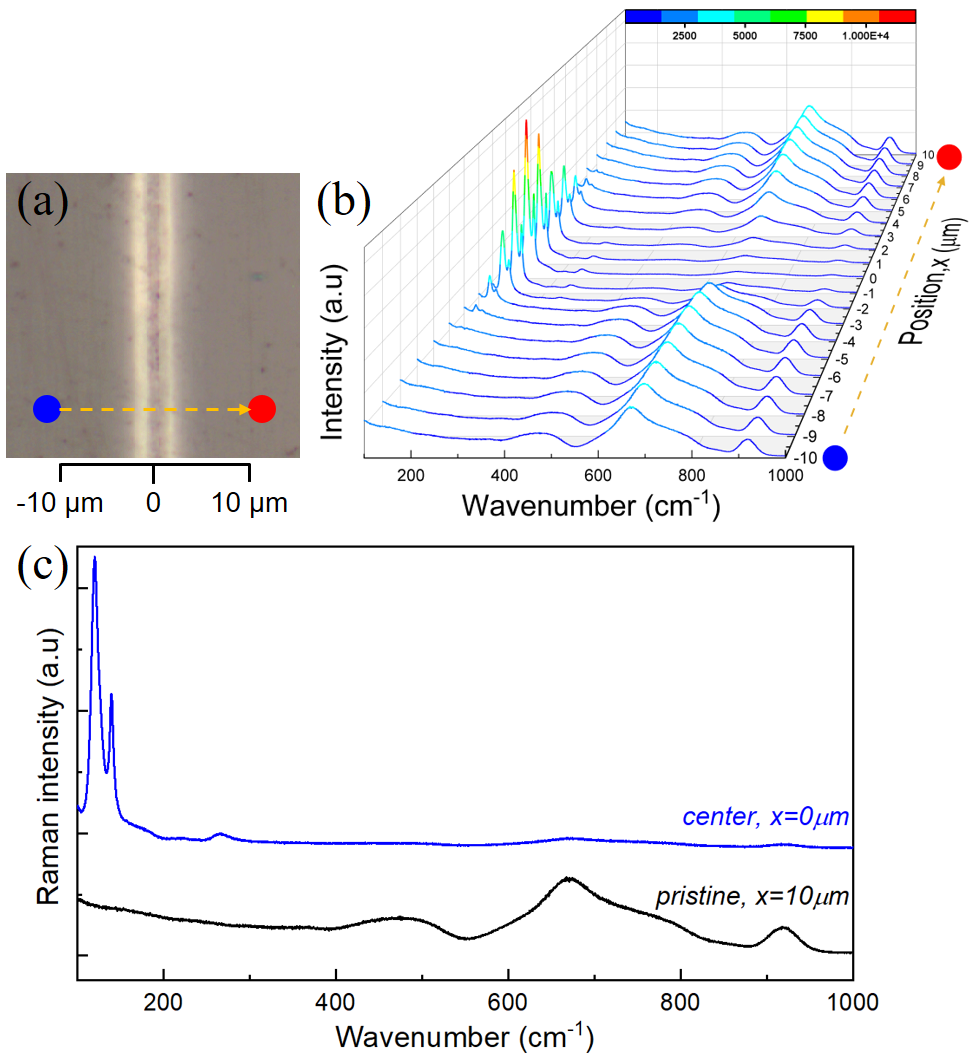}
\caption{\label{fig:wide}Raman spectra of femtosecond laser inscribed lines. (a) Optical images of potassium-tungsten tellurite surface after femtosecond laser irradiation with laser parameters of 10 nJ, 0.5 mm/s (13 J/mm\textsuperscript{2}). (b) Raman spectra collected by scanning laser across the modified zone and spectra collected at each 1 $\mu$m. Yellow dash line indicates scanned line with Raman laser. (c) Raman spectra taken at the center of modification at \textit{x=0 $\mu$m} and in a unmodified region at a position of \textit{x=10 $\mu$m} away from the center.}
\end{figure}

To investigate further the laser-induced modifications, we performed Raman spectroscopy measurements on various specimens. Figure 5 and Figure 6 show the optical microscopy images of potassium-tungsten tellurite surfaces after femtosecond laser irradiation and the corresponding Raman spectra collected at points distributed across the laser modified zone with a spatial sampling period of 1 $\mu$m. The morphology of the line in Figure 5 is similar to the one in Figure 2(a). Gradual changes in the short range order of the structure are observed along the scanned line as shown in Figure 5(b). The spectrum at the beginning of the Raman scan (at \textit{x=-10 $\mu$m}) is similar to the one of the pristine glass, shown in Figure 5(c). The peak assignment, presented in Table 1, was performed based on the work from Kosuge \textit{et al.} \cite{kosuge_thermal_1998}. The main peaks are found around 490, 610, 670, 720, 790, 860 and 920 cm\textsuperscript{-1}.
Along the scanned line in Figure 5(b), a few peaks with lower wavenumbers start to appear three micrometers away from the center \textit{(x=0 $\mu$m)}. As their intensity gradually increases towards the center, the main peaks assigned to structures in the glass vanish in Figure 5(c). Raman spectra are deconvoluted into Gaussian bands and accordingly, the peaks at 119, 139, 170, 220, 264 and 355 cm\textsuperscript{-1} are observed at the center of modified region (peak assignment is performed based on \cite{torrie_raman_1970, yannopoulos_structure_2020, ananth_kumar_scrutiny_2017, yuan_compositional_2012, song_superlong_2008, salmon-gamboa_vibrational_2018, sekiya_structural_1994, UPENDER2010468}). Peaks at 119, 139, 220 and 260 cm\textsuperscript{-1} belong to a crystallized trigonal tellurium (t-Te) phase. 170 cm\textsuperscript{-1} is attributed to Te-Te homopolar bonds in amorphous tellurium (a-Te). As reported by others, Te-Te homopolar bonds are not stable and crystallize above 285 K, however, they can be found up to 473 K \cite{cheng_femtosecond_2018, yannopoulos_structure_2020, ananth_kumar_scrutiny_2017}. This result indicates the existence of a chain-like structure within the amorphous phase after femtosecond pulses irradiation. In addition to the new peaks appearance mentioned above, a weak peak at 360 cm\textsuperscript{-1} is found within the nanostructures. This peak is attributed to the bending vibration mode of W–O–W in WO\textsubscript{6} octahedra \cite{sekiya_structural_1994, UPENDER2010468}. Note that the microstructure at 100 kHz in Figure 1(h). shows no crystallization along the modified region (see Supplementary Material for Raman spectra of the microstructure in Fig. 1(h).).

Figure 6 shows the relative intensities of the main peaks at various fluences. From outside of the modification to its boundary, Te-O bonds concentration seems to decay in quantity. At the boundary of nanostructures, there is sharp increase in t-Te/glassy-TeO\textsubscript{2} ratio, followed by a decrease towards the center. There, the t-Te content decreases with laser fluence and transforms to glassy-TeO\textsubscript{2}, the intensity of peaks related to TeO\textsubscript{3}+TeO\textsubscript{3+1} increases, and finally, TeO\textsubscript{4} trigonal bipyramids (tbp) decreases as reported for the tellurite glass with higher glass-modifier content \cite{sekiya_structural_1994}. 

Furthermore, the crystallization event expands beyond the focal volume and nanostructures, as in Figure 6. At fixed translation velocities, an increase in pulse energy significantly affects the crystallization width. Above the non-linear absorption threshold, the crystallized width increases approximately by 2 $\mu$m for every 10 nJ when the translation velocity is fixed to 0.5 mm/s. Likewise, at a fixed pulse energy, the translation velocity has a dominant effect on the crystallized region. In Figure 6(c), self-organization is localized within 10 $\mu$m, whereas crystallized zone are found in a wider region (up to 40 $\mu$m). These observations are consistent with a temperature-driven process.
\begin{table*}
\centering
\caption{\label{tab:wide}
Peak assignments of Raman spectra of laser modified lines with references.}
\begin{ruledtabular}
\begin{tabular}{ l p{14cm} }
Wavenumber(cm\textsuperscript{-1}) & Assignment\\
\hline
119 & Symmetric stretching vibrations of crystalline Te (A1 mode) \\
139 & Doubly degenerated vibration modes of crystalline Te (E\textsubscript{TO} mode) \\
176 & Te-Te homopolar bonds in amorphous Te \\
220 & Doubly degenerated vibration modes of crystalline Te (E\textsubscript{TO} mode)\\
260 & Second-order E vibrational mode of crystalline Te\\
356 & Bending vibrations of W–O–W in WO\textsubscript{6} octahedra\\
490 & Symmetrical stretching vibration mode of Te-O-Te linkages\\
610 & Vibration mode of continuous network composed of TeO\textsubscript{4} trigonal bipyramid (tbp)\\
670 & Anti-symmetric vibration mode of Te-O-Te linkages by two nonequivalent Te-O bonds in TeO\textsubscript{4} tbp \\
720 & Stretching vibrations mode between Te and non-bridging oxygen (NBO) of TeO\textsubscript{3+1} (distorted tbp) polyhedra and TeO\textsubscript{3} trigonal pyramid (tp) \\
790 & Te-O\textsuperscript{-} stretching vibration mode of TeO\textsubscript{3+1} polyhedra \\
860 & W-O stretching vibration mode in WO\textsubscript{4} or WO\textsubscript{6} units \\
920 & Stretching vibration mode of the W-O\textsuperscript{-} and W=O terminal bonds associated with WO\textsubscript{4} or WO\textsubscript{6} polyhedra \\
\end{tabular}
\end{ruledtabular}
\end{table*}

\begin{figure*}
\includegraphics[width=15cm]{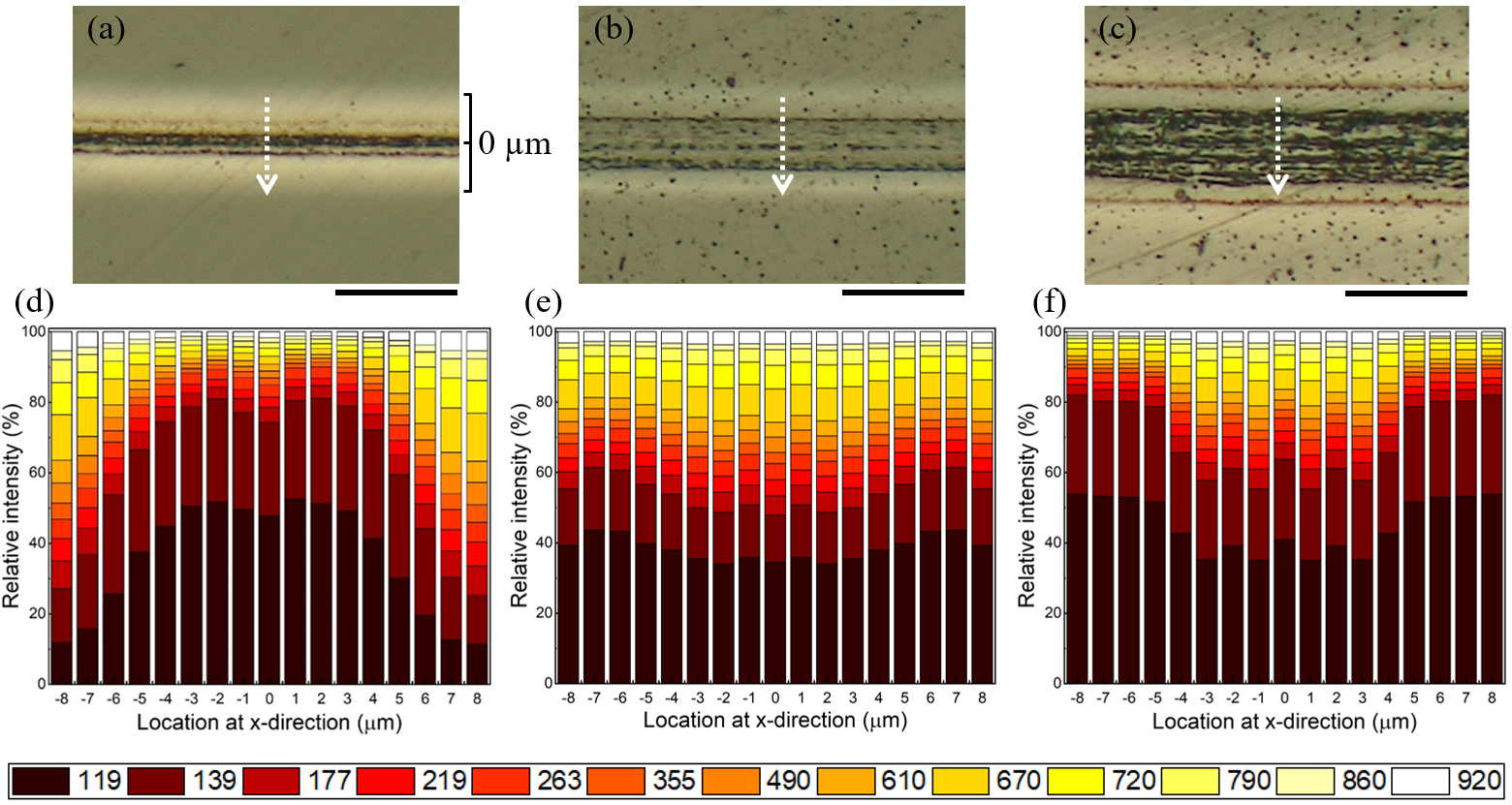}
\caption{\label{fig:wide} Raman spectra of femtosecond laser inscribed lines. Optical microscopy images of potassium-tungsten tellurite surface after femtosecond laser irradiation at 1 MHz: (a) 20 nJ, 0.5 mm/s (26 J/mm\textsuperscript{2}); (b) 200 nJ, 10 mm/s (13 J/mm\textsuperscript{2}) and (c) 200 nJ, 0.5 mm/s (262 J/mm\textsuperscript{2}). (d-f) Relative intensities of the main Raman spectra peaks found in modified zones (by scanning across the modified zone with 1 $\mu$m of step size). White dash lines indicate the Raman scanned lines and directions. The color-coded legend shows the main peaks and their assigned colors. Black scale bar is 20 $\mu$m.}
\end{figure*}

Both, Raman spectroscopy and EDS results of the self-organized nanostructures are consistent with one another. In brief, after the femtosecond laser irradiation, a cohort of tungsten atoms migrate to the vicinity of the nanostructures, whereas oxygen is removed along modified zones. A possible structural change is as follows. The lone electron pair in Te impedes the short-range atomic arrangement in TeO\textsubscript{2}, making the first coordination sphere around Te highly asymmetric and variable. TeO\textsubscript{3} trigonal pyramids (tp) and TeO\textsubscript{3+1} distorted trigonal bi-pyramids (tbp) proportions are greatly reduced and change appreciably in the spatially confined medium, as Te becomes depleted by deoxygenation. In the meantime, tungsten coordination number might be altered, inducing change in the Raman peak-intensity ratio I\textsubscript{355}/I\textsubscript{920}. Previous study investigated the formation of a-Te from t-Te upon femtosecond laser irradiation \cite{cheng_femtosecond_2018}. In their observations, amorphization takes places after the first laser pulse, while crystallization occurs during subsequent pulses. Elemental dissociation of Te has also been reported for ZnTe \cite{li_pressure-induced_2010, kshirsagar_photoinduced_2013}. In our case, as we observe the end product only, we do not know the transformation pathways starting from the initial tellurite glass that ultimately led to the formation of a-Te or t-Te phases as we are observing here. We can only speculate on a possible sequence of phase transformation. In such scenario, a-Te form during the first few laser pulses as a result of elemental redistribution and thermal melting, subsequently followed by re-crystallization after further exposure. The observations of chain-like a-Te at 170 cm\textsuperscript{-1} in the Raman spectra support the formation of the amorphous phase. In addition, defective states in the tellurium created by deoxygenation may contribute to the formation of these chain-like structures at high temperatures. At lower pulse energy, the modified zone consists of a-Te and t-Te. At higher pulse energy, the final microstructure depends on the exposure dose (i.e. the number of cumulative pulses). For low-pulse overlapping ratio, only t-Te and a-Te are found after a few pulses exposure. As pulses accumulate, the temperature becomes high enough over longer time to recrystallize immediately t-Te that had turned into an amorphous state. During this process, defects and dangling bonds of Te may react with neighboring oxygen molecules. This oxidation process in the nanostructured zone support the following phase transformation sequence: t-Te $\rightarrow$ a-Te $\rightarrow$ glassy-TeO\textsubscript{2} \cite{vasileiadis_photo-induced_2014}. Localized ionization of the ambient air due to the beam high-peak power ($\sim$10-30 TW/cm\textsuperscript{2}) \cite{PRL} in the vicinity of the material surface is likely to further enhance this effect by feeding ionized oxygen atoms to the oxidation process. In parallel, the migration of tungsten elements contributes to preserving the glassy-TeO\textsubscript{2} along the nanoplanes. 

Self-organized nanoplanes in the center of laser affected zones and secondary nanogratings at the boundaries have notable differences in term of structures highlighted in Figure 5 and Figure 6. Similar to Raman results, EDS shows slight differences in elemental content (Fig. 4). We attribute the formation of these secondary gratings to mechanical rupture and cracks formation during solidification caused by a difference of thermal expansion coefficient at the boundary between modified and unmodified materials. To support this observation, we note that since W\textsuperscript{6+} is a heavy metal ion, the depletion of W from the matrix in the nanogratings distort the glass network and contributes to high stress levels at the phase boundaries, eventually leading to cracks formation. The periodicity of these boundary nanocracks originates from the periodicity of the nanogratings phase itself. The fact that the periodicity of the secondary nanogratings does not change with laser fluence, further supports that this process is of thermo-mechanical origin, and not a immediate outcome of laser-exposure. Note that some of these nanocracks are also seldom present in the middle of the laser affected zones, in between nanoplanes (see Fig. 1 for a few examples), which is consequent with a fracture during solidification model, as proposed above. It also explains why fewer to none are found in the case of a polarization perpendicular to the writing direction as the nanoplanes vertical morphology leads to fewer stress-concentration points, prone to crack-nucleation. 

The fabrication of ultra-thin crystalline tellurium films on the surface of TeO\textsubscript{2}-based glass provides also a means for exploring how laser-induced geometric confinement in two dimensions can cause changes at the short- and medium- range structural order in a non-crystalline medium. Analyzing the Raman spectra, it is found that the structure of glassy-TeO\textsubscript{2} undergoes a dramatic modification. These results may enable direct-write tailored physical properties, e.g. the Te/TeO\textsubscript{2} interface has shown photo-conduction in the UV region \cite{OISHI198729}.

\begin{figure*}[ht!]
\includegraphics[width=16cm]{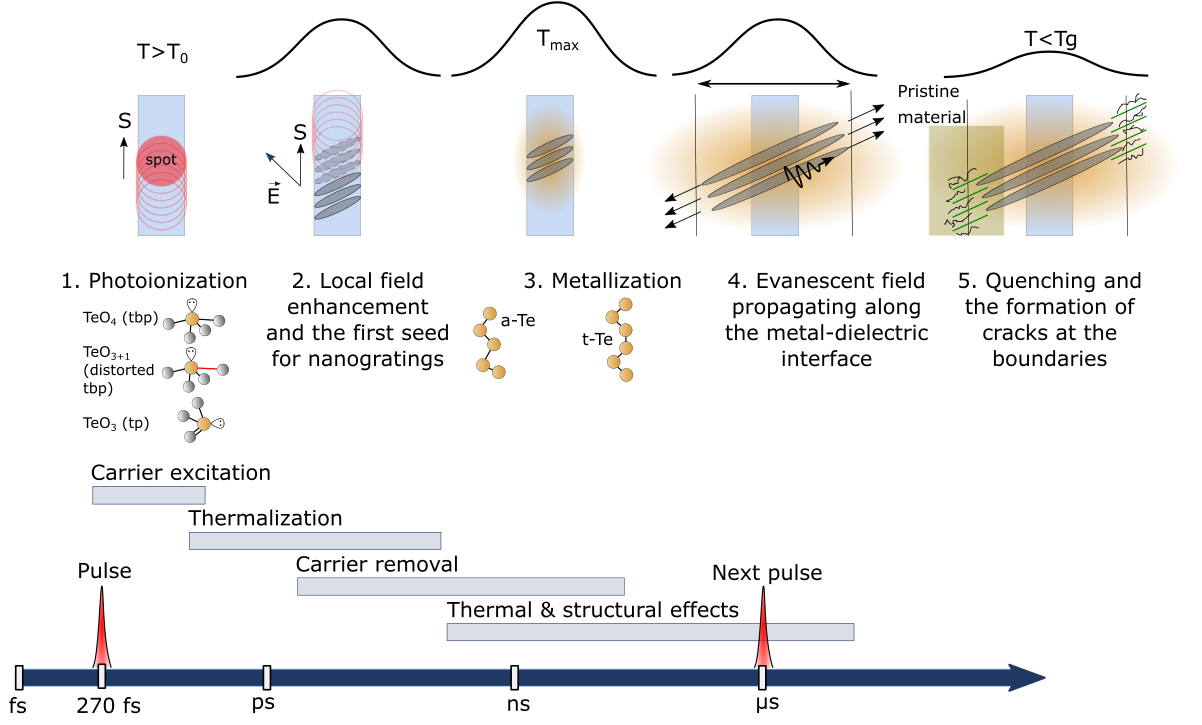}
\caption{\label{fig:wide}Phenomenological interpretation for the sequential formation of self-organized nanostructures beyond the focal volume and its related metallization on ternary-tellurite glass surface: \textit{Step 1.} Ionization through multi-photon absorption, \textit{step 2.} Formation of nanogratings by local field enhancement after multi-pulse irradiation, \textit{step 3.} Metallization by elemental decomposition and deoxygenation followed by thermal melting, \textit{step 4.} Evanescent wave formation and SPP propagation at the metal/dielectric interface and finally, \textit{step 5.} Ultrafast cooling and the formation of the cracks due to thermal expansion mismatch between modified and parent glass.}
\end{figure*}

\subsection{Proposed scenario to explain the various phenomena}
We propose the following interpretation that we illustrate schematically in Figure 7 and Figure 8. 

Let us first examine the steps leading towards self-organized structures that span beyond the focal volume and examine the key events occurring during laser exposure according to our phenomenological model.

\textit{Step 1:} During femtosecond laser inscription, multi-photon ionization seeds first electrons that further lead, through avalanche ionization, to the build-up of an under-dense plasma. As bonds are broken within the main structural units of the glass, weakly bonded ions, such as O\textsuperscript{2-} and Te\textsuperscript{4+} form. 

\textit{Step 2:} The first few pulses create an electric-field enhancement through the formation of surface plasmon-polariton pairs (SPPs) that initiate the formation of self-organized nanostructures within the focal volume \cite{bonse_role_2009, huang2014}. For specific conditions, surface electromagnetic waves (SEWs) through surface plasmon-polariton coupling (SPPs) can be excited. SPPs may originate from delocalized coherent electron density oscillations and propagate along the interfaces between the two different material phases emerging locally during the self-organization process. Driven by the laser electromagnetic field, SPPs propagate and dissipate away from both sides of the laser-exposed zone with a preferential direction defined by the self-organized nanoplanes orientation. For the excitation of SPPs, specific conditions for the dielectric permittivity of the involved media have to be fulfilled. Particularly, for irradiation with ultrashort laser pulses, this excitation channel can be enabled even for semiconductors and dielectrics, as the initially non-plasmonic material can transiently be turned into a metallic state, enabling SPPs once a critical density of electrons in the conduction band is exceeded. The merge of SPPs leads to a spatial modulation of the local energy field distribution, which, through absorption mechanism, is imprinted in the specimen. Note that SPPs formation has been showed theoretically and experimentally on silicon upon femtosecond laser irradiation \cite{sokolowski_2010}.

\textit{Step 3:} As the pulses accumulate, the temperature builds up gradually. Thermal melting and elemental decomposition with deoxygenation events stimulate the crystallization of tellurium. As the molten state of modified tellurite glass includes charged entities (modifier cations, such as W\textsuperscript{6+}, and Te\textsuperscript{4+}) as well as neutral polar molecules such as TeO\textsubscript{2}, metallization occurs through intra-/inter- molecular charges exchanges \cite{marini_high-pressure_2012}. Under strong electrical field, the molten material becomes polarized and forms elongated clusters as a result of the balance between electrical forces on induced surface charges and interfacial tension forces. The retention of local electrical neutrality has structurally an effect on the connection and coordination of the polyhedra. As observed with the Raman spectra, migration of tungsten results in the decrease of Te-O-Te linkages and W=O bonds and the formation of W-O-Te linkages, similar to WO\textsubscript{3}-rich tellurite glass \cite{sekiya_structural_1994}. Thus, the elemental redistribution is followed by deoxygenation and metallization process.

\textit{Step 4:} As heat builds up locally, a thin layer of crystalline Te forms. Here, we make the hypothesis that an evanescent SPP field propagates at the interface. As a dipole moment is created, forces between polarized particles arise and keep them apart. This mechanism would explain the periodicity of the nanostructures observed \textit{outside} the zone under direct-laser exposure. To support this model, we note that similar ordering induced by an evanescent field has been observed in small dielectric particles where the periodicity depends on the particle diameters and the ionic polarizability that affects the repulsive/attractive potential between the charged particles \cite{mellor_array_2006}.

\textit{Step 5:} Depending on the pulse overlapping ratio, the temperature in the focal volume may stay high, long enough, to allow for crystalline thin layers to grow. During cooling, due to the thermal expansion coefficients mismatch between modified and parent material phases, high stress concentration builds up at the interface and causes the formation of interstitial micro-cracks localized in between nanoplanes, and notably, where stress-concentration occurs at the boundary between pristine and modified materials. 

\begin{figure}
\includegraphics[width=8.5cm]{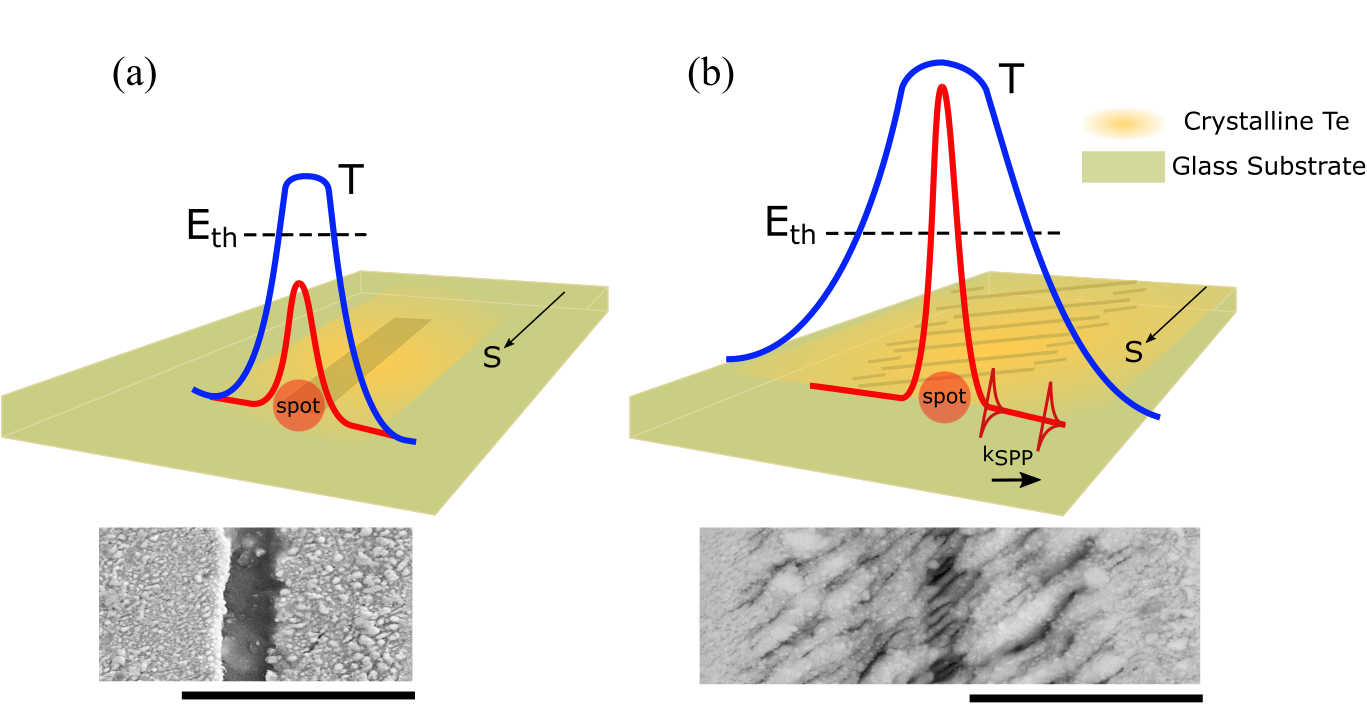}
\caption{\label{fig:wide}Phenomenological interpretation of the wider nanostructures formation on potassium-tungsten tellurite glass. When multiple pulses accumulate, the temperature rise causes the material under exposure to melt. Despite the thermal-cumulative regime leading to the bulk heating of the material, nanostructures and secondary nanostructures form at higher pulse energy, but are not found for lower pulse energy. Temperature gradient indicated as blue and field strength illustrated as red line on modified zone (a) 10 nJ, 0.5 mm/s (13 J/mm\textsuperscript{2}); (b) 200 nJ, 0.5 mm/s (262 J/mm\textsuperscript{2}) (both at 1 MHz).The scale bars in both images are 5 $\mu$m.}
\end{figure}

Let us now discuss the effect of field strength as observed in Figure 8.
Our interpretation is as follows. The laser intense electrical field determines the field-strength atoms are exposed. It will affect the condition for SPPs to exist \cite{bruevich} as it will impact the plasma electron density. In such model, below a given threshold (here found around 20 nJ), the field-intensity is not high enough to sustain the self-organization mechanism described before. Note that a similar mechanism was proposed by Liao \textit{et al.} \cite{Liao2015}.

In summary, the dynamics of formation of this structure is complex. The structure under exposure is not static and evolve dynamically as the materials heat up and as nanostructures gradually form. Our interpretation is based on final state of matter, out of which we reconstruct a possible scenario. As ‘metallic seeds’ form through elemental decomposition at high temperature, the main hypothesis is that they start promoting SPPs due at their interface when subsequent femtosecond laser pulse arrives. In this quasi-static evolution scenario, we assume that after each pulse, the system can be considered as a ‘metastable’ configuration for the next incoming pulse, which supposes that viscosity remains sufficiently high, so that this approximation is valid, otherwise, the in-homogeneity (and phase separation) could not be maintained from pulse-to-pulse. From the material phase evidences collected, we observe the formation of this tellurium phase surrounding the glassy phase itself localized within the nanoplanes. The growth of these nanoplanes outside the laser focus is explained by localized field-enhancement. Here, we have considered the hypothetical presence of SPPs to explain the field-enhancement that certainly also involves quasi-cylindrical waves that may play a constructive role in this field-enhancement mechanism, adding to the effect of the SPPs through cross-conversion mechanism \cite{Lalanne2009}. To confirm this scenario, pump-probe experiments or alike could be considered for identifying the onset and sequence of intermediate events proposed here.
 
\section{CONCLUSIONS}

We reported on a complex mechanism of self-organization induced by femtosecond lasers at the surface of potassium-tungsten tellurite glass. Specifically, we observed the formation of self-organized nanostructures oriented perpendicular to the laser polarization that span beyond the focal volume, and this, in a regime where heat accumulates. Furthermore, our observations indicate a localized laser-induced metallization occuring under femtosecond laser irradiation and combined with a self-organization process. The formation of nanoplanes are accompanied by elemental redistribution event that results in localized crystallization. These results are promising for nonlinear optics and photonic applications in which high refractive index, high polarizability and dielectric constants play an important role. Adding the possibility of controlling the formation of crystalline tellurium phases in the dielectric matrix give a further degree of freedom in the engineering of composite photonic structures, in particular, considering that tellurium has plasmonic-like properties.

\begin{acknowledgments}

The Galatea Lab is thankful to the sponsorship of Richemont International and to Tokyo Tech (TIT) for providing the tellurite glass. The authors are thankful to Amplitude Systèmes for providing the experimental laser sources used in this study.

In this paper, G.T. wrote the draft paper, G.T. performed the most part of the experiments presented in this paper, T.K. did the first exposure lines and preliminary observation, G.T. and Y.B. interpreted the experimental results, and analyzed the experimental data. T.K. produced the bulk material and measurement data of various bulk properties. Y.B. designed and supervised the research. All the authors discussed and revised the paper. 

\end{acknowledgments}

\bibliography{Bibliography}

\begin{thebibliography}{56}%
\makeatletter
\providecommand \@ifxundefined [1]{%
 \@ifx{#1\undefined}
}%
\providecommand \@ifnum [1]{%
 \ifnum #1\expandafter \@firstoftwo
 \else \expandafter \@secondoftwo
 \fi
}%
\providecommand \@ifx [1]{%
 \ifx #1\expandafter \@firstoftwo
 \else \expandafter \@secondoftwo
 \fi
}%
\providecommand \natexlab [1]{#1}%
\providecommand \enquote  [1]{``#1''}%
\providecommand \bibnamefont  [1]{#1}%
\providecommand \bibfnamefont [1]{#1}%
\providecommand \citenamefont [1]{#1}%
\providecommand \href@noop [0]{\@secondoftwo}%
\providecommand \href [0]{\begingroup \@sanitize@url \@href}%
\providecommand \@href[1]{\@@startlink{#1}\@@href}%
\providecommand \@@href[1]{\endgroup#1\@@endlink}%
\providecommand \@sanitize@url [0]{\catcode `\\12\catcode `\$12\catcode
  `\&12\catcode `\#12\catcode `\^12\catcode `\_12\catcode `\%12\relax}%
\providecommand \@@startlink[1]{}%
\providecommand \@@endlink[0]{}%
\providecommand \url  [0]{\begingroup\@sanitize@url \@url }%
\providecommand \@url [1]{\endgroup\@href {#1}{\urlprefix }}%
\providecommand \urlprefix  [0]{URL }%
\providecommand \Eprint [0]{\href }%
\providecommand \doibase [0]{https://doi.org/}%
\providecommand \selectlanguage [0]{\@gobble}%
\providecommand \bibinfo  [0]{\@secondoftwo}%
\providecommand \bibfield  [0]{\@secondoftwo}%
\providecommand \translation [1]{[#1]}%
\providecommand \BibitemOpen [0]{}%
\providecommand \bibitemStop [0]{}%
\providecommand \bibitemNoStop [0]{.\EOS\space}%
\providecommand \EOS [0]{\spacefactor3000\relax}%
\providecommand \BibitemShut  [1]{\csname bibitem#1\endcsname}%
\let\auto@bib@innerbib\@empty
\bibitem [{\citenamefont {Bonse}\ \emph {et~al.}(2012)\citenamefont {Bonse},
  \citenamefont {Kr{\"u}ger}, \citenamefont {H{\"o}hm},\ and\ \citenamefont
  {Rosenfeld}}]{bonse_femtosecond_2012}%
  \BibitemOpen
  \bibfield  {author} {\bibinfo {author} {\bibfnamefont {J.}~\bibnamefont
  {Bonse}}, \bibinfo {author} {\bibfnamefont {J.}~\bibnamefont {Kr{\"u}ger}},
  \bibinfo {author} {\bibfnamefont {S.}~\bibnamefont {H{\"o}hm}},\ and\
  \bibinfo {author} {\bibfnamefont {A.}~\bibnamefont {Rosenfeld}},\ }\href
  {https://doi.org/10.2351/1.4712658} {\bibfield  {journal} {\bibinfo
  {journal} {Journal of Laser Applications}\ }\textbf {\bibinfo {volume}
  {24}},\ \bibinfo {pages} {042006} (\bibinfo {year} {2012})}\BibitemShut
  {NoStop}%
\bibitem [{\citenamefont {Vorobyev}\ and\ \citenamefont
  {Guo}(2008)}]{vorobyev_colorizing_2008}%
  \BibitemOpen
  \bibfield  {author} {\bibinfo {author} {\bibfnamefont {A.~Y.}\ \bibnamefont
  {Vorobyev}}\ and\ \bibinfo {author} {\bibfnamefont {C.}~\bibnamefont {Guo}},\
  }\href@noop {} {\bibfield  {journal} {\bibinfo  {journal} {Appl. Phys.
  Lett.}\ }\textbf {\bibinfo {volume} {92}},\ \bibinfo {pages} {4} (\bibinfo
  {year} {2008})}\BibitemShut {NoStop}%
\bibitem [{\citenamefont {Baldacchini}\ \emph {et~al.}(2006)\citenamefont
  {Baldacchini}, \citenamefont {Carey}, \citenamefont {Zhou},\ and\
  \citenamefont {Mazur}}]{baldacchini_superhydrophobic_2006}%
  \BibitemOpen
  \bibfield  {author} {\bibinfo {author} {\bibfnamefont {T.}~\bibnamefont
  {Baldacchini}}, \bibinfo {author} {\bibfnamefont {J.~E.}\ \bibnamefont
  {Carey}}, \bibinfo {author} {\bibfnamefont {M.}~\bibnamefont {Zhou}},\ and\
  \bibinfo {author} {\bibfnamefont {E.}~\bibnamefont {Mazur}},\ }\href
  {https://doi.org/10.1021/la053374k} {\bibfield  {journal} {\bibinfo
  {journal} {Langmuir}\ }\textbf {\bibinfo {volume} {22}},\ \bibinfo {pages}
  {4917} (\bibinfo {year} {2006})}\BibitemShut {NoStop}%
\bibitem [{\citenamefont {Eichst{\"a}dt}\ \emph {et~al.}(2011)\citenamefont
  {Eichst{\"a}dt}, \citenamefont {R{\"o}mer},\ and\ \citenamefont {in’t
  Veld}}]{eichstadt}%
  \BibitemOpen
  \bibfield  {author} {\bibinfo {author} {\bibfnamefont {J.}~\bibnamefont
  {Eichst{\"a}dt}}, \bibinfo {author} {\bibfnamefont {G.}~\bibnamefont
  {R{\"o}mer}},\ and\ \bibinfo {author} {\bibfnamefont {A.~H.}\ \bibnamefont
  {in’t Veld}},\ }\href
  {https://doi.org/https://doi.org/10.1016/j.phpro.2011.03.099} {\bibfield
  {journal} {\bibinfo  {journal} {Physics Procedia}\ }\textbf {\bibinfo
  {volume} {12}},\ \bibinfo {pages} {7 } (\bibinfo {year} {2011})},\ \bibinfo
  {note} {lasers in Manufacturing 2011 - Proceedings of the Sixth International
  WLT Conference on Lasers in Manufacturing}\BibitemShut {NoStop}%
\bibitem [{\citenamefont {Bonse}\ and\ \citenamefont
  {Gr{\"a}f}(2020)}]{bonse2020}%
  \BibitemOpen
  \bibfield  {author} {\bibinfo {author} {\bibfnamefont {J.}~\bibnamefont
  {Bonse}}\ and\ \bibinfo {author} {\bibfnamefont {S.}~\bibnamefont
  {Gr{\"a}f}},\ }\href {https://doi.org/https://doi.org/10.1002/lpor.202000215}
  {\bibfield  {journal} {\bibinfo  {journal} {Laser \& Photonics Reviews}\
  }\textbf {\bibinfo {volume} {14}},\ \bibinfo {pages} {2000215} (\bibinfo
  {year} {2020})}\BibitemShut {NoStop}%
\bibitem [{\citenamefont {Wu}\ \emph {et~al.}(2003)\citenamefont {Wu},
  \citenamefont {Ma}, \citenamefont {Fang}, \citenamefont {Liao}, \citenamefont
  {Yu}, \citenamefont {Chen},\ and\ \citenamefont
  {Wang}}]{wu_femtosecond_2003}%
  \BibitemOpen
  \bibfield  {author} {\bibinfo {author} {\bibfnamefont {Q.}~\bibnamefont
  {Wu}}, \bibinfo {author} {\bibfnamefont {Y.}~\bibnamefont {Ma}}, \bibinfo
  {author} {\bibfnamefont {R.}~\bibnamefont {Fang}}, \bibinfo {author}
  {\bibfnamefont {Y.}~\bibnamefont {Liao}}, \bibinfo {author} {\bibfnamefont
  {Q.}~\bibnamefont {Yu}}, \bibinfo {author} {\bibfnamefont {X.}~\bibnamefont
  {Chen}},\ and\ \bibinfo {author} {\bibfnamefont {K.}~\bibnamefont {Wang}},\
  }\href {https://doi.org/10.1063/1.1561581} {\bibfield  {journal} {\bibinfo
  {journal} {Applied Physics Letters}\ }\textbf {\bibinfo {volume} {82}},\
  \bibinfo {pages} {1703} (\bibinfo {year} {2003})}\BibitemShut {NoStop}%
\bibitem [{\citenamefont {Borowiec}\ and\ \citenamefont
  {Haugen}(2003)}]{borowiec_subwavelength_2003}%
  \BibitemOpen
  \bibfield  {author} {\bibinfo {author} {\bibfnamefont {A.}~\bibnamefont
  {Borowiec}}\ and\ \bibinfo {author} {\bibfnamefont {H.~K.}\ \bibnamefont
  {Haugen}},\ }\href {https://doi.org/10.1063/1.1586457} {\bibfield  {journal}
  {\bibinfo  {journal} {Applied Physics Letters}\ }\textbf {\bibinfo {volume}
  {82}},\ \bibinfo {pages} {4462} (\bibinfo {year} {2003})}\BibitemShut
  {NoStop}%
\bibitem [{\citenamefont {Jia}\ \emph {et~al.}(2005)\citenamefont {Jia},
  \citenamefont {Chen}, \citenamefont {Huang}, \citenamefont {Zhao},
  \citenamefont {Qiu}, \citenamefont {Li}, \citenamefont {Xu}, \citenamefont
  {He}, \citenamefont {Zhang},\ and\ \citenamefont
  {Kuroda}}]{jia_formation_2005}%
  \BibitemOpen
  \bibfield  {author} {\bibinfo {author} {\bibfnamefont {T.~Q.}\ \bibnamefont
  {Jia}}, \bibinfo {author} {\bibfnamefont {H.~X.}\ \bibnamefont {Chen}},
  \bibinfo {author} {\bibfnamefont {M.}~\bibnamefont {Huang}}, \bibinfo
  {author} {\bibfnamefont {F.~L.}\ \bibnamefont {Zhao}}, \bibinfo {author}
  {\bibfnamefont {J.~R.}\ \bibnamefont {Qiu}}, \bibinfo {author} {\bibfnamefont
  {R.~X.}\ \bibnamefont {Li}}, \bibinfo {author} {\bibfnamefont {Z.~Z.}\
  \bibnamefont {Xu}}, \bibinfo {author} {\bibfnamefont {X.~K.}\ \bibnamefont
  {He}}, \bibinfo {author} {\bibfnamefont {J.}~\bibnamefont {Zhang}},\ and\
  \bibinfo {author} {\bibfnamefont {H.}~\bibnamefont {Kuroda}},\ }\href
  {https://doi.org/10.1103/PhysRevB.72.125429} {\bibfield  {journal} {\bibinfo
  {journal} {Physical Review B}\ }\textbf {\bibinfo {volume} {72}},\ \bibinfo
  {pages} {125429} (\bibinfo {year} {2005})}\BibitemShut {NoStop}%
\bibitem [{\citenamefont {Bonse}\ \emph {et~al.}(2005)\citenamefont {Bonse},
  \citenamefont {Munz},\ and\ \citenamefont {Sturm}}]{bonse_structure_2005}%
  \BibitemOpen
  \bibfield  {author} {\bibinfo {author} {\bibfnamefont {J.}~\bibnamefont
  {Bonse}}, \bibinfo {author} {\bibfnamefont {M.}~\bibnamefont {Munz}},\ and\
  \bibinfo {author} {\bibfnamefont {H.}~\bibnamefont {Sturm}},\ }\href
  {https://doi.org/10.1063/1.1827919} {\bibfield  {journal} {\bibinfo
  {journal} {Journal of Applied Physics}\ }\textbf {\bibinfo {volume} {97}},\
  \bibinfo {pages} {013538} (\bibinfo {year} {2005})}\BibitemShut {NoStop}%
\bibitem [{\citenamefont {Martsinovski\u{\i}}\ \emph
  {et~al.}(2008)\citenamefont {Martsinovski\u{\i}}, \citenamefont {Shandybina},
  \citenamefont {Smirnov}, \citenamefont {Zabotnov}, \citenamefont
  {Golovan’}, \citenamefont {Timoshenko},\ and\ \citenamefont
  {Kashkarov}}]{martsinovskii_ultrashort_2008}%
  \BibitemOpen
  \bibfield  {author} {\bibinfo {author} {\bibfnamefont {G.~A.}\ \bibnamefont
  {Martsinovski\u{\i}}}, \bibinfo {author} {\bibfnamefont {G.~D.}\ \bibnamefont
  {Shandybina}}, \bibinfo {author} {\bibfnamefont {D.~S.}\ \bibnamefont
  {Smirnov}}, \bibinfo {author} {\bibfnamefont {S.~V.}\ \bibnamefont
  {Zabotnov}}, \bibinfo {author} {\bibfnamefont {L.~A.}\ \bibnamefont
  {Golovan’}}, \bibinfo {author} {\bibfnamefont {V.~Y.}\ \bibnamefont
  {Timoshenko}},\ and\ \bibinfo {author} {\bibfnamefont {P.~K.}\ \bibnamefont
  {Kashkarov}},\ }\href {https://doi.org/10.1134/S0030400X08070114} {\bibfield
  {journal} {\bibinfo  {journal} {Optics and Spectroscopy}\ }\textbf {\bibinfo
  {volume} {105}},\ \bibinfo {pages} {67} (\bibinfo {year} {2008})}\BibitemShut
  {NoStop}%
\bibitem [{\citenamefont {Miyaji}\ and\ \citenamefont
  {Miyazaki}(2008)}]{miyaji_origin_2008}%
  \BibitemOpen
  \bibfield  {author} {\bibinfo {author} {\bibfnamefont {G.}~\bibnamefont
  {Miyaji}}\ and\ \bibinfo {author} {\bibfnamefont {K.}~\bibnamefont
  {Miyazaki}},\ }\href {https://doi.org/10.1364/OE.16.016265} {\bibfield
  {journal} {\bibinfo  {journal} {Optics Express}\ }\textbf {\bibinfo {volume}
  {16}},\ \bibinfo {pages} {16265} (\bibinfo {year} {2008})}\BibitemShut
  {NoStop}%
\bibitem [{\citenamefont {Shugaev}\ \emph {et~al.}(2017)\citenamefont
  {Shugaev}, \citenamefont {Gnilitskyi}, \citenamefont {Bulgakova},\ and\
  \citenamefont {Zhigilei}}]{shugaev_mechanism_2017}%
  \BibitemOpen
  \bibfield  {author} {\bibinfo {author} {\bibfnamefont {M.~V.}\ \bibnamefont
  {Shugaev}}, \bibinfo {author} {\bibfnamefont {I.}~\bibnamefont {Gnilitskyi}},
  \bibinfo {author} {\bibfnamefont {N.~M.}\ \bibnamefont {Bulgakova}},\ and\
  \bibinfo {author} {\bibfnamefont {L.~V.}\ \bibnamefont {Zhigilei}},\ }\href
  {https://doi.org/10.1103/PhysRevB.96.205429} {\bibfield  {journal} {\bibinfo
  {journal} {Physical Review B}\ }\textbf {\bibinfo {volume} {96}},\ \bibinfo
  {pages} {205429} (\bibinfo {year} {2017})}\BibitemShut {NoStop}%
\bibitem [{\citenamefont {Reif}\ \emph {et~al.}(2002)\citenamefont {Reif},
  \citenamefont {Costache}, \citenamefont {Henyk},\ and\ \citenamefont
  {Pandelov}}]{reif_ripples_2002}%
  \BibitemOpen
  \bibfield  {author} {\bibinfo {author} {\bibfnamefont {J.}~\bibnamefont
  {Reif}}, \bibinfo {author} {\bibfnamefont {F.}~\bibnamefont {Costache}},
  \bibinfo {author} {\bibfnamefont {M.}~\bibnamefont {Henyk}},\ and\ \bibinfo
  {author} {\bibfnamefont {S.~V.}\ \bibnamefont {Pandelov}},\ }\href
  {https://doi.org/10.1016/S0169-4332(02)00450-6} {\bibfield  {journal}
  {\bibinfo  {journal} {Applied Surface Science}\ }\textbf {\bibinfo {volume}
  {197-198}},\ \bibinfo {pages} {891} (\bibinfo {year} {2002})}\BibitemShut
  {NoStop}%
\bibitem [{\citenamefont {Li}\ \emph {et~al.}(2014)\citenamefont {Li},
  \citenamefont {Zhang}, \citenamefont {Li}, \citenamefont {Dai}, \citenamefont
  {Lan},\ and\ \citenamefont {Tie}}]{li_formation_2014}%
  \BibitemOpen
  \bibfield  {author} {\bibinfo {author} {\bibfnamefont {X.-F.}\ \bibnamefont
  {Li}}, \bibinfo {author} {\bibfnamefont {C.-Y.}\ \bibnamefont {Zhang}},
  \bibinfo {author} {\bibfnamefont {H.}~\bibnamefont {Li}}, \bibinfo {author}
  {\bibfnamefont {Q.-F.}\ \bibnamefont {Dai}}, \bibinfo {author} {\bibfnamefont
  {S.}~\bibnamefont {Lan}},\ and\ \bibinfo {author} {\bibfnamefont {S.-L.}\
  \bibnamefont {Tie}},\ }\href {https://doi.org/10.1364/OE.22.028086}
  {\bibfield  {journal} {\bibinfo  {journal} {Optics Express}\ }\textbf
  {\bibinfo {volume} {22}},\ \bibinfo {pages} {28086} (\bibinfo {year}
  {2014})}\BibitemShut {NoStop}%
\bibitem [{\citenamefont {Costache}\ \emph {et~al.}(2004)\citenamefont
  {Costache}, \citenamefont {Kouteva-Arguirova},\ and\ \citenamefont
  {Reif}}]{costache_subdamagethreshold_2004}%
  \BibitemOpen
  \bibfield  {author} {\bibinfo {author} {\bibfnamefont {F.}~\bibnamefont
  {Costache}}, \bibinfo {author} {\bibfnamefont {S.}~\bibnamefont
  {Kouteva-Arguirova}},\ and\ \bibinfo {author} {\bibfnamefont
  {J.}~\bibnamefont {Reif}},\ }\href
  {https://doi.org/10.1007/s00339-004-2803-y} {\bibfield  {journal} {\bibinfo
  {journal} {Applied Physics A}\ }\textbf {\bibinfo {volume} {79}},\ \bibinfo
  {pages} {1429} (\bibinfo {year} {2004})}\BibitemShut {NoStop}%
\bibitem [{\citenamefont {Gr{\"a}f}\ \emph {et~al.}(2017)\citenamefont
  {Gr{\"a}f}, \citenamefont {Kunz},\ and\ \citenamefont
  {M{\"u}ller}}]{graf_formation_2017}%
  \BibitemOpen
  \bibfield  {author} {\bibinfo {author} {\bibfnamefont {S.}~\bibnamefont
  {Gr{\"a}f}}, \bibinfo {author} {\bibfnamefont {C.}~\bibnamefont {Kunz}},\
  and\ \bibinfo {author} {\bibfnamefont {F.}~\bibnamefont {M{\"u}ller}},\
  }\href {https://doi.org/10.3390/ma10080933} {\bibfield  {journal} {\bibinfo
  {journal} {Materials}\ }\textbf {\bibinfo {volume} {10}},\ \bibinfo {pages}
  {933} (\bibinfo {year} {2017})}\BibitemShut {NoStop}%
\bibitem [{\citenamefont {Ding}\ \emph {et~al.}(2009)\citenamefont {Ding},
  \citenamefont {Cancado}, \citenamefont {Novotny}, \citenamefont {Knox},
  \citenamefont {Anderson}, \citenamefont {Jani}, \citenamefont {Linhardt},
  \citenamefont {Blackwell},\ and\ \citenamefont
  {K{\"u}nzler}}]{ding_micro-raman_2009}%
  \BibitemOpen
  \bibfield  {author} {\bibinfo {author} {\bibfnamefont {L.}~\bibnamefont
  {Ding}}, \bibinfo {author} {\bibfnamefont {L.~G.}\ \bibnamefont {Cancado}},
  \bibinfo {author} {\bibfnamefont {L.}~\bibnamefont {Novotny}}, \bibinfo
  {author} {\bibfnamefont {W.~H.}\ \bibnamefont {Knox}}, \bibinfo {author}
  {\bibfnamefont {N.}~\bibnamefont {Anderson}}, \bibinfo {author}
  {\bibfnamefont {D.}~\bibnamefont {Jani}}, \bibinfo {author} {\bibfnamefont
  {J.}~\bibnamefont {Linhardt}}, \bibinfo {author} {\bibfnamefont {R.~I.}\
  \bibnamefont {Blackwell}},\ and\ \bibinfo {author} {\bibfnamefont {J.~F.}\
  \bibnamefont {K{\"u}nzler}},\ }\href
  {https://doi.org/10.1364/JOSAB.26.000595} {\bibfield  {journal} {\bibinfo
  {journal} {Journal of the Optical Society of America B}\ }\textbf {\bibinfo
  {volume} {26}},\ \bibinfo {pages} {595} (\bibinfo {year} {2009})}\BibitemShut
  {NoStop}%
\bibitem [{\citenamefont {Ly}\ \emph {et~al.}(2017)\citenamefont {Ly},
  \citenamefont {Shen}, \citenamefont {Negres}, \citenamefont {Carr},
  \citenamefont {Alessi}, \citenamefont {Bude}, \citenamefont {Rigatti},\ and\
  \citenamefont {Laurence}}]{ly_role_2017}%
  \BibitemOpen
  \bibfield  {author} {\bibinfo {author} {\bibfnamefont {S.}~\bibnamefont
  {Ly}}, \bibinfo {author} {\bibfnamefont {N.}~\bibnamefont {Shen}}, \bibinfo
  {author} {\bibfnamefont {R.~A.}\ \bibnamefont {Negres}}, \bibinfo {author}
  {\bibfnamefont {C.~W.}\ \bibnamefont {Carr}}, \bibinfo {author}
  {\bibfnamefont {D.~A.}\ \bibnamefont {Alessi}}, \bibinfo {author}
  {\bibfnamefont {J.~D.}\ \bibnamefont {Bude}}, \bibinfo {author}
  {\bibfnamefont {A.}~\bibnamefont {Rigatti}},\ and\ \bibinfo {author}
  {\bibfnamefont {T.~A.}\ \bibnamefont {Laurence}},\ }\href
  {https://doi.org/10.1364/OE.25.015161} {\bibfield  {journal} {\bibinfo
  {journal} {Optics Express}\ }\textbf {\bibinfo {volume} {25}},\ \bibinfo
  {pages} {15161} (\bibinfo {year} {2017})}\BibitemShut {NoStop}%
\bibitem [{\citenamefont {Wuttig}\ and\ \citenamefont
  {Yamada}(2007)}]{wuttig_phase-change_2007}%
  \BibitemOpen
  \bibfield  {author} {\bibinfo {author} {\bibfnamefont {M.}~\bibnamefont
  {Wuttig}}\ and\ \bibinfo {author} {\bibfnamefont {N.}~\bibnamefont
  {Yamada}},\ }\href {https://doi.org/10.1038/nmat2009} {\bibfield  {journal}
  {\bibinfo  {journal} {Nature Materials}\ }\textbf {\bibinfo {volume} {6}},\
  \bibinfo {pages} {824} (\bibinfo {year} {2007})}\BibitemShut {NoStop}%
\bibitem [{\citenamefont {Matsunaga}\ \emph {et~al.}(2011)\citenamefont
  {Matsunaga}, \citenamefont {Akola}, \citenamefont {Kohara}, \citenamefont
  {Honma}, \citenamefont {Kobayashi}, \citenamefont {Ikenaga}, \citenamefont
  {Jones}, \citenamefont {Yamada}, \citenamefont {Takata},\ and\ \citenamefont
  {Kojima}}]{matsunaga_local_2011}%
  \BibitemOpen
  \bibfield  {author} {\bibinfo {author} {\bibfnamefont {T.}~\bibnamefont
  {Matsunaga}}, \bibinfo {author} {\bibfnamefont {J.}~\bibnamefont {Akola}},
  \bibinfo {author} {\bibfnamefont {S.}~\bibnamefont {Kohara}}, \bibinfo
  {author} {\bibfnamefont {T.}~\bibnamefont {Honma}}, \bibinfo {author}
  {\bibfnamefont {K.}~\bibnamefont {Kobayashi}}, \bibinfo {author}
  {\bibfnamefont {E.}~\bibnamefont {Ikenaga}}, \bibinfo {author} {\bibfnamefont
  {R.~O.}\ \bibnamefont {Jones}}, \bibinfo {author} {\bibfnamefont
  {N.}~\bibnamefont {Yamada}}, \bibinfo {author} {\bibfnamefont
  {M.}~\bibnamefont {Takata}},\ and\ \bibinfo {author} {\bibfnamefont
  {R.}~\bibnamefont {Kojima}},\ }\href {https://doi.org/10.1038/nmat2931}
  {\bibfield  {journal} {\bibinfo  {journal} {Nature Materials}\ }\textbf
  {\bibinfo {volume} {10}},\ \bibinfo {pages} {129} (\bibinfo {year}
  {2011})}\BibitemShut {NoStop}%
\bibitem [{\citenamefont {Li}\ \emph {et~al.}(2010)\citenamefont {Li},
  \citenamefont {Wang}, \citenamefont {Li}, \citenamefont {Ma}, \citenamefont
  {Cui},\ and\ \citenamefont {Zou}}]{li_pressure-induced_2010}%
  \BibitemOpen
  \bibfield  {author} {\bibinfo {author} {\bibfnamefont {Z.}~\bibnamefont
  {Li}}, \bibinfo {author} {\bibfnamefont {H.}~\bibnamefont {Wang}}, \bibinfo
  {author} {\bibfnamefont {Y.}~\bibnamefont {Li}}, \bibinfo {author}
  {\bibfnamefont {Y.}~\bibnamefont {Ma}}, \bibinfo {author} {\bibfnamefont
  {T.}~\bibnamefont {Cui}},\ and\ \bibinfo {author} {\bibfnamefont
  {G.}~\bibnamefont {Zou}},\ }\href
  {https://doi.org/10.1088/1367-2630/12/4/043058} {\bibfield  {journal}
  {\bibinfo  {journal} {New Journal of Physics}\ }\textbf {\bibinfo {volume}
  {12}},\ \bibinfo {pages} {043058} (\bibinfo {year} {2010})}\BibitemShut
  {NoStop}%
\bibitem [{\citenamefont {Çelikbilek}\ \emph {et~al.}(2011)\citenamefont
  {Çelikbilek}, \citenamefont {Ersundu}, \citenamefont {Solak},\ and\
  \citenamefont {Aydin}}]{celikbilek_crystallization_2011}%
  \BibitemOpen
  \bibfield  {author} {\bibinfo {author} {\bibfnamefont {M.}~\bibnamefont
  {Çelikbilek}}, \bibinfo {author} {\bibfnamefont {A.}~\bibnamefont
  {Ersundu}}, \bibinfo {author} {\bibfnamefont {N.}~\bibnamefont {Solak}},\
  and\ \bibinfo {author} {\bibfnamefont {S.}~\bibnamefont {Aydin}},\ }\href
  {https://doi.org/10.1016/j.jnoncrysol.2010.09.012} {\bibfield  {journal}
  {\bibinfo  {journal} {Journal of Non-Crystalline Solids}\ }\textbf {\bibinfo
  {volume} {357}},\ \bibinfo {pages} {88} (\bibinfo {year} {2011})}\BibitemShut
  {NoStop}%
\bibitem [{\citenamefont {El-Mallawany}(2011)}]{el2011tellurite}%
  \BibitemOpen
  \bibfield  {author} {\bibinfo {author} {\bibfnamefont {R.}~\bibnamefont
  {El-Mallawany}},\ }\href {https://books.google.ch/books?id=rEJhNu7XNcsC}
  {\emph {\bibinfo {title} {Tellurite Glasses Handbook: Physical Properties and
  Data, Second Edition}}}\ (\bibinfo  {publisher} {Taylor \& Francis},\
  \bibinfo {year} {2011})\BibitemShut {NoStop}%
\bibitem [{\citenamefont {Kosuge}\ \emph {et~al.}(1998)\citenamefont {Kosuge},
  \citenamefont {Benino}, \citenamefont {Dimitrov}, \citenamefont {Sato},\ and\
  \citenamefont {Komatsu}}]{kosuge_thermal_1998}%
  \BibitemOpen
  \bibfield  {author} {\bibinfo {author} {\bibfnamefont {T.}~\bibnamefont
  {Kosuge}}, \bibinfo {author} {\bibfnamefont {Y.}~\bibnamefont {Benino}},
  \bibinfo {author} {\bibfnamefont {V.}~\bibnamefont {Dimitrov}}, \bibinfo
  {author} {\bibfnamefont {R.}~\bibnamefont {Sato}},\ and\ \bibinfo {author}
  {\bibfnamefont {T.}~\bibnamefont {Komatsu}},\ }\href
  {https://doi.org/10.1016/S0022-3093(98)00800-X} {\bibfield  {journal}
  {\bibinfo  {journal} {Journal of Non-Crystalline Solids}\ }\textbf {\bibinfo
  {volume} {242}},\ \bibinfo {pages} {154} (\bibinfo {year}
  {1998})}\BibitemShut {NoStop}%
\bibitem [{\citenamefont {Çelikbilek Ersundu}\ \emph
  {et~al.}(2017)\citenamefont {Çelikbilek Ersundu}, \citenamefont {Ersundu},
  \citenamefont {Sayyed}, \citenamefont {Lakshminarayana},\ and\ \citenamefont
  {Aydin}}]{celikbilek_ersundu_evaluation_2017}%
  \BibitemOpen
  \bibfield  {author} {\bibinfo {author} {\bibfnamefont {M.}~\bibnamefont
  {Çelikbilek Ersundu}}, \bibinfo {author} {\bibfnamefont {A.}~\bibnamefont
  {Ersundu}}, \bibinfo {author} {\bibfnamefont {M.}~\bibnamefont {Sayyed}},
  \bibinfo {author} {\bibfnamefont {G.}~\bibnamefont {Lakshminarayana}},\ and\
  \bibinfo {author} {\bibfnamefont {S.}~\bibnamefont {Aydin}},\ }\href
  {https://doi.org/10.1016/j.jallcom.2017.04.223} {\bibfield  {journal}
  {\bibinfo  {journal} {Journal of Alloys and Compounds}\ }\textbf {\bibinfo
  {volume} {714}},\ \bibinfo {pages} {278} (\bibinfo {year}
  {2017})}\BibitemShut {NoStop}%
\bibitem [{\citenamefont {Sidkey}\ and\ \citenamefont
  {Gaafar}(2004)}]{sidkey_ultrasonic_2004}%
  \BibitemOpen
  \bibfield  {author} {\bibinfo {author} {\bibfnamefont {M.}~\bibnamefont
  {Sidkey}}\ and\ \bibinfo {author} {\bibfnamefont {M.}~\bibnamefont
  {Gaafar}},\ }\href {https://doi.org/10.1016/j.physb.2003.11.005} {\bibfield
  {journal} {\bibinfo  {journal} {Physica B: Condensed Matter}\ }\textbf
  {\bibinfo {volume} {348}},\ \bibinfo {pages} {46} (\bibinfo {year}
  {2004})}\BibitemShut {NoStop}%
\bibitem [{\citenamefont {Tanabe}\ \emph {et~al.}(1990)\citenamefont {Tanabe},
  \citenamefont {Hirao},\ and\ \citenamefont
  {Soga}}]{tanabe_upconversion_nodate}%
  \BibitemOpen
  \bibfield  {author} {\bibinfo {author} {\bibfnamefont {S.}~\bibnamefont
  {Tanabe}}, \bibinfo {author} {\bibfnamefont {K.}~\bibnamefont {Hirao}},\ and\
  \bibinfo {author} {\bibfnamefont {N.}~\bibnamefont {Soga}},\ }\href
  {https://doi.org/https://doi.org/10.1016/0022-3093(90)90228-E} {\bibfield
  {journal} {\bibinfo  {journal} {Journal of Non-Crystalline Solids}\ }\textbf
  {\bibinfo {volume} {122}},\ \bibinfo {pages} {79 } (\bibinfo {year}
  {1990})}\BibitemShut {NoStop}%
\bibitem [{\citenamefont {Wang}\ \emph {et~al.}(1994)\citenamefont {Wang},
  \citenamefont {Vogel},\ and\ \citenamefont {Snitzer}}]{wang1194}%
  \BibitemOpen
  \bibfield  {author} {\bibinfo {author} {\bibfnamefont {J.}~\bibnamefont
  {Wang}}, \bibinfo {author} {\bibfnamefont {E.}~\bibnamefont {Vogel}},\ and\
  \bibinfo {author} {\bibfnamefont {E.}~\bibnamefont {Snitzer}},\ }\href
  {https://doi.org/https://doi.org/10.1016/0925-3467(94)90004-3} {\bibfield
  {journal} {\bibinfo  {journal} {Optical Materials}\ }\textbf {\bibinfo
  {volume} {3}},\ \bibinfo {pages} {187 } (\bibinfo {year} {1994})}\BibitemShut
  {NoStop}%
\bibitem [{\citenamefont {Nasu}\ \emph {et~al.}(1990)\citenamefont {Nasu},
  \citenamefont {Matsushita}, \citenamefont {Kamiya}, \citenamefont
  {Kobayashi},\ and\ \citenamefont {Kubodera}}]{nasu1990}%
  \BibitemOpen
  \bibfield  {author} {\bibinfo {author} {\bibfnamefont {H.}~\bibnamefont
  {Nasu}}, \bibinfo {author} {\bibfnamefont {O.}~\bibnamefont {Matsushita}},
  \bibinfo {author} {\bibfnamefont {K.}~\bibnamefont {Kamiya}}, \bibinfo
  {author} {\bibfnamefont {H.}~\bibnamefont {Kobayashi}},\ and\ \bibinfo
  {author} {\bibfnamefont {K.}~\bibnamefont {Kubodera}},\ }\href
  {https://doi.org/https://doi.org/10.1016/0022-3093(90)90274-P} {\bibfield
  {journal} {\bibinfo  {journal} {Journal of Non-Crystalline Solids}\ }\textbf
  {\bibinfo {volume} {124}},\ \bibinfo {pages} {275 } (\bibinfo {year}
  {1990})}\BibitemShut {NoStop}%
\bibitem [{\citenamefont {Wu}\ \emph {et~al.}(2019)\citenamefont {Wu},
  \citenamefont {Huang}, \citenamefont {Wang}, \citenamefont {Zhao},
  \citenamefont {Ma}, \citenamefont {Xiang}, \citenamefont {Li}, \citenamefont
  {Ponraj}, \citenamefont {Dhanabalan},\ and\ \citenamefont {Zhang}}]{wu2019}%
  \BibitemOpen
  \bibfield  {author} {\bibinfo {author} {\bibfnamefont {L.}~\bibnamefont
  {Wu}}, \bibinfo {author} {\bibfnamefont {W.}~\bibnamefont {Huang}}, \bibinfo
  {author} {\bibfnamefont {Y.}~\bibnamefont {Wang}}, \bibinfo {author}
  {\bibfnamefont {J.}~\bibnamefont {Zhao}}, \bibinfo {author} {\bibfnamefont
  {D.}~\bibnamefont {Ma}}, \bibinfo {author} {\bibfnamefont {Y.}~\bibnamefont
  {Xiang}}, \bibinfo {author} {\bibfnamefont {J.}~\bibnamefont {Li}}, \bibinfo
  {author} {\bibfnamefont {J.~S.}\ \bibnamefont {Ponraj}}, \bibinfo {author}
  {\bibfnamefont {S.~C.}\ \bibnamefont {Dhanabalan}},\ and\ \bibinfo {author}
  {\bibfnamefont {H.}~\bibnamefont {Zhang}},\ }\href
  {https://doi.org/10.1002/adfm.201806346} {\bibfield  {journal} {\bibinfo
  {journal} {Advanced Functional Materials}\ }\textbf {\bibinfo {volume}
  {29}},\ \bibinfo {pages} {1806346} (\bibinfo {year} {2019})}\BibitemShut
  {NoStop}%
\bibitem [{\citenamefont {Rajesh}\ and\ \citenamefont
  {Bellouard}(2010)}]{Rajesh}%
  \BibitemOpen
  \bibfield  {author} {\bibinfo {author} {\bibfnamefont {S.}~\bibnamefont
  {Rajesh}}\ and\ \bibinfo {author} {\bibfnamefont {Y.}~\bibnamefont
  {Bellouard}},\ }\href
  {http://www.opticsexpress.org/abstract.cfm?URI=oe-18-20-21490} {\bibfield
  {journal} {\bibinfo  {journal} {Opt. Express}\ }\textbf {\bibinfo {volume}
  {18}},\ \bibinfo {pages} {21490} (\bibinfo {year} {2010})}\BibitemShut
  {NoStop}%
\bibitem [{\citenamefont {Luedtke}\ \emph {et~al.}(2011)\citenamefont
  {Luedtke}, \citenamefont {Gao},\ and\ \citenamefont
  {Landman}}]{luedtke_dielectric_2011}%
  \BibitemOpen
  \bibfield  {author} {\bibinfo {author} {\bibfnamefont {W.~D.}\ \bibnamefont
  {Luedtke}}, \bibinfo {author} {\bibfnamefont {J.}~\bibnamefont {Gao}},\ and\
  \bibinfo {author} {\bibfnamefont {U.}~\bibnamefont {Landman}},\ }\href
  {https://doi.org/10.1021/jp206673j} {\bibfield  {journal} {\bibinfo
  {journal} {The Journal of Physical Chemistry C}\ }\textbf {\bibinfo {volume}
  {115}},\ \bibinfo {pages} {20343} (\bibinfo {year} {2011})}\BibitemShut
  {NoStop}%
\bibitem [{\citenamefont {Bonse}\ \emph {et~al.}(2017)\citenamefont {Bonse},
  \citenamefont {Hohm}, \citenamefont {Kirner}, \citenamefont {Rosenfeld},\
  and\ \citenamefont {Kruger}}]{bonse_laser-induced_2017}%
  \BibitemOpen
  \bibfield  {author} {\bibinfo {author} {\bibfnamefont {J.}~\bibnamefont
  {Bonse}}, \bibinfo {author} {\bibfnamefont {S.}~\bibnamefont {Hohm}},
  \bibinfo {author} {\bibfnamefont {S.~V.}\ \bibnamefont {Kirner}}, \bibinfo
  {author} {\bibfnamefont {A.}~\bibnamefont {Rosenfeld}},\ and\ \bibinfo
  {author} {\bibfnamefont {J.}~\bibnamefont {Kruger}},\ }\href
  {https://doi.org/10.1109/JSTQE.2016.2614183} {\bibfield  {journal} {\bibinfo
  {journal} {IEEE Journal of Selected Topics in Quantum Electronics}\ }\textbf
  {\bibinfo {volume} {23}},\ \bibinfo {pages} {7581030} (\bibinfo {year}
  {2017})}\BibitemShut {NoStop}%
\bibitem [{\citenamefont {Pamler}\ and\ \citenamefont
  {Marinero}(1987)}]{pamler_transient_1987}%
  \BibitemOpen
  \bibfield  {author} {\bibinfo {author} {\bibfnamefont {W.}~\bibnamefont
  {Pamler}}\ and\ \bibinfo {author} {\bibfnamefont {E.~E.}\ \bibnamefont
  {Marinero}},\ }\href {https://doi.org/10.1063/1.337939} {\bibfield  {journal}
  {\bibinfo  {journal} {Journal of Applied Physics}\ }\textbf {\bibinfo
  {volume} {61}},\ \bibinfo {pages} {2294} (\bibinfo {year}
  {1987})}\BibitemShut {NoStop}%
\bibitem [{\citenamefont {Cheng}\ \emph {et~al.}(2018)\citenamefont {Cheng},
  \citenamefont {Teitelbaum}, \citenamefont {Gao},\ and\ \citenamefont
  {Nelson}}]{cheng_femtosecond_2018}%
  \BibitemOpen
  \bibfield  {author} {\bibinfo {author} {\bibfnamefont {Y.-H.}\ \bibnamefont
  {Cheng}}, \bibinfo {author} {\bibfnamefont {S.~W.}\ \bibnamefont
  {Teitelbaum}}, \bibinfo {author} {\bibfnamefont {F.~Y.}\ \bibnamefont
  {Gao}},\ and\ \bibinfo {author} {\bibfnamefont {K.~A.}\ \bibnamefont
  {Nelson}},\ }\href {https://doi.org/10.1103/PhysRevB.98.134112} {\bibfield
  {journal} {\bibinfo  {journal} {Physical Review B}\ }\textbf {\bibinfo
  {volume} {98}},\ \bibinfo {pages} {134112} (\bibinfo {year}
  {2018})}\BibitemShut {NoStop}%
\bibitem [{\citenamefont {Akagi}\ \emph {et~al.}(1999)\citenamefont {Akagi},
  \citenamefont {Handa}, \citenamefont {Ohtori}, \citenamefont {Hannon},
  \citenamefont {Tatsumisago},\ and\ \citenamefont {Umesaki}}]{Akagi_1999}%
  \BibitemOpen
  \bibfield  {author} {\bibinfo {author} {\bibfnamefont {R.}~\bibnamefont
  {Akagi}}, \bibinfo {author} {\bibfnamefont {K.}~\bibnamefont {Handa}},
  \bibinfo {author} {\bibfnamefont {N.}~\bibnamefont {Ohtori}}, \bibinfo
  {author} {\bibfnamefont {A.~C.}\ \bibnamefont {Hannon}}, \bibinfo {author}
  {\bibfnamefont {M.}~\bibnamefont {Tatsumisago}},\ and\ \bibinfo {author}
  {\bibfnamefont {N.}~\bibnamefont {Umesaki}},\ }\href
  {https://doi.org/10.7567/jjaps.38s1.160} {\bibfield  {journal} {\bibinfo
  {journal} {Japanese Journal of Applied Physics}\ }\textbf {\bibinfo {volume}
  {38}},\ \bibinfo {pages} {160} (\bibinfo {year} {1999})}\BibitemShut
  {NoStop}%
\bibitem [{\citenamefont {Torrie}(1970)}]{torrie_raman_1970}%
  \BibitemOpen
  \bibfield  {author} {\bibinfo {author} {\bibfnamefont {B.}~\bibnamefont
  {Torrie}},\ }\href {https://doi.org/10.1016/0038-1098(70)90343-1} {\bibfield
  {journal} {\bibinfo  {journal} {Solid State Communications}\ }\textbf
  {\bibinfo {volume} {8}},\ \bibinfo {pages} {1899} (\bibinfo {year}
  {1970})}\BibitemShut {NoStop}%
\bibitem [{\citenamefont {Yannopoulos}(2020)}]{yannopoulos_structure_2020}%
  \BibitemOpen
  \bibfield  {author} {\bibinfo {author} {\bibfnamefont {S.~N.}\ \bibnamefont
  {Yannopoulos}},\ }\href {https://doi.org/10.1007/s10854-020-03310-0}
  {\bibfield  {journal} {\bibinfo  {journal} {Journal of Materials Science:
  Materials in Electronics}\ }\textbf {\bibinfo {volume} {31}},\ \bibinfo
  {pages} {7565} (\bibinfo {year} {2020})}\BibitemShut {NoStop}%
\bibitem [{\citenamefont {Ananth~Kumar}\ \emph {et~al.}(2017)\citenamefont
  {Ananth~Kumar}, \citenamefont {Mousa}, \citenamefont {Chithra~Lekha},
  \citenamefont {Mahmoud},\ and\ \citenamefont
  {Qamhieh}}]{ananth_kumar_scrutiny_2017}%
  \BibitemOpen
  \bibfield  {author} {\bibinfo {author} {\bibfnamefont {R.~T.}\ \bibnamefont
  {Ananth~Kumar}}, \bibinfo {author} {\bibfnamefont {H.~A.}\ \bibnamefont
  {Mousa}}, \bibinfo {author} {\bibfnamefont {P.}~\bibnamefont
  {Chithra~Lekha}}, \bibinfo {author} {\bibfnamefont {S.~T.}\ \bibnamefont
  {Mahmoud}},\ and\ \bibinfo {author} {\bibfnamefont {N.}~\bibnamefont
  {Qamhieh}},\ }\href {https://doi.org/10.1088/1742-6596/869/1/012018}
  {\bibfield  {journal} {\bibinfo  {journal} {Journal of Physics: Conference
  Series}\ }\textbf {\bibinfo {volume} {869}},\ \bibinfo {pages} {012018}
  (\bibinfo {year} {2017})}\BibitemShut {NoStop}%
\bibitem [{\citenamefont {Yuan}\ \emph {et~al.}(2012)\citenamefont {Yuan},
  \citenamefont {Yang}, \citenamefont {Chen}, \citenamefont {Qian},
  \citenamefont {Shen}, \citenamefont {Zhang},\ and\ \citenamefont
  {Jiang}}]{yuan_compositional_2012}%
  \BibitemOpen
  \bibfield  {author} {\bibinfo {author} {\bibfnamefont {J.}~\bibnamefont
  {Yuan}}, \bibinfo {author} {\bibfnamefont {Q.}~\bibnamefont {Yang}}, \bibinfo
  {author} {\bibfnamefont {D.~D.}\ \bibnamefont {Chen}}, \bibinfo {author}
  {\bibfnamefont {Q.}~\bibnamefont {Qian}}, \bibinfo {author} {\bibfnamefont
  {S.~X.}\ \bibnamefont {Shen}}, \bibinfo {author} {\bibfnamefont {Q.~Y.}\
  \bibnamefont {Zhang}},\ and\ \bibinfo {author} {\bibfnamefont {Z.~H.}\
  \bibnamefont {Jiang}},\ }\href {https://doi.org/10.1063/1.4717980} {\bibfield
   {journal} {\bibinfo  {journal} {Journal of Applied Physics}\ }\textbf
  {\bibinfo {volume} {111}},\ \bibinfo {pages} {103511} (\bibinfo {year}
  {2012})}\BibitemShut {NoStop}%
\bibitem [{\citenamefont {Song}\ \emph {et~al.}(2008)\citenamefont {Song},
  \citenamefont {Lin}, \citenamefont {Zhan}, \citenamefont {Tian},
  \citenamefont {Liu},\ and\ \citenamefont {Yu}}]{song_superlong_2008}%
  \BibitemOpen
  \bibfield  {author} {\bibinfo {author} {\bibfnamefont {J.-M.}\ \bibnamefont
  {Song}}, \bibinfo {author} {\bibfnamefont {Y.-Z.}\ \bibnamefont {Lin}},
  \bibinfo {author} {\bibfnamefont {Y.-J.}\ \bibnamefont {Zhan}}, \bibinfo
  {author} {\bibfnamefont {Y.-C.}\ \bibnamefont {Tian}}, \bibinfo {author}
  {\bibfnamefont {G.}~\bibnamefont {Liu}},\ and\ \bibinfo {author}
  {\bibfnamefont {S.-H.}\ \bibnamefont {Yu}},\ }\href
  {https://doi.org/10.1021/cg701125k} {\bibfield  {journal} {\bibinfo
  {journal} {Crystal Growth \& Design}\ }\textbf {\bibinfo {volume} {8}},\
  \bibinfo {pages} {1902} (\bibinfo {year} {2008})}\BibitemShut {NoStop}%
\bibitem [{\citenamefont {Salmón-Gamboa}\ \emph {et~al.}(2018)\citenamefont
  {Salmón-Gamboa}, \citenamefont {Barajas-Aguilar}, \citenamefont
  {Ruiz-Ortega}, \citenamefont {Garay-Tapia},\ and\ \citenamefont
  {Jiménez-Sandoval}}]{salmon-gamboa_vibrational_2018}%
  \BibitemOpen
  \bibfield  {author} {\bibinfo {author} {\bibfnamefont {J.~U.}\ \bibnamefont
  {Salmón-Gamboa}}, \bibinfo {author} {\bibfnamefont {A.~H.}\ \bibnamefont
  {Barajas-Aguilar}}, \bibinfo {author} {\bibfnamefont {L.~I.}\ \bibnamefont
  {Ruiz-Ortega}}, \bibinfo {author} {\bibfnamefont {A.~M.}\ \bibnamefont
  {Garay-Tapia}},\ and\ \bibinfo {author} {\bibfnamefont {S.~J.}\ \bibnamefont
  {Jiménez-Sandoval}},\ }\href {https://doi.org/10.1038/s41598-018-26461-x}
  {\bibfield  {journal} {\bibinfo  {journal} {Scientific Reports}\ }\textbf
  {\bibinfo {volume} {8}},\ \bibinfo {pages} {8093} (\bibinfo {year}
  {2018})}\BibitemShut {NoStop}%
\bibitem [{\citenamefont {Sekiya}\ \emph {et~al.}(1994)\citenamefont {Sekiya},
  \citenamefont {Mochida},\ and\ \citenamefont
  {Ogawa}}]{sekiya_structural_1994}%
  \BibitemOpen
  \bibfield  {author} {\bibinfo {author} {\bibfnamefont {T.}~\bibnamefont
  {Sekiya}}, \bibinfo {author} {\bibfnamefont {N.}~\bibnamefont {Mochida}},\
  and\ \bibinfo {author} {\bibfnamefont {S.}~\bibnamefont {Ogawa}},\ }\href
  {https://doi.org/10.1016/0022-3093(94)90067-1} {\bibfield  {journal}
  {\bibinfo  {journal} {Journal of Non-Crystalline Solids}\ }\textbf {\bibinfo
  {volume} {176}},\ \bibinfo {pages} {105} (\bibinfo {year}
  {1994})}\BibitemShut {NoStop}%
\bibitem [{\citenamefont {Upender}\ \emph {et~al.}(2010)\citenamefont
  {Upender}, \citenamefont {Ramesh}, \citenamefont {Prasad}, \citenamefont
  {Sathe},\ and\ \citenamefont {Mouli}}]{UPENDER2010468}%
  \BibitemOpen
  \bibfield  {author} {\bibinfo {author} {\bibfnamefont {G.}~\bibnamefont
  {Upender}}, \bibinfo {author} {\bibfnamefont {S.}~\bibnamefont {Ramesh}},
  \bibinfo {author} {\bibfnamefont {M.}~\bibnamefont {Prasad}}, \bibinfo
  {author} {\bibfnamefont {V.}~\bibnamefont {Sathe}},\ and\ \bibinfo {author}
  {\bibfnamefont {V.}~\bibnamefont {Mouli}},\ }\href
  {https://doi.org/https://doi.org/10.1016/j.jallcom.2010.06.006} {\bibfield
  {journal} {\bibinfo  {journal} {Journal of Alloys and Compounds}\ }\textbf
  {\bibinfo {volume} {504}},\ \bibinfo {pages} {468 } (\bibinfo {year}
  {2010})}\BibitemShut {NoStop}%
\bibitem [{\citenamefont {Kshirsagar}\ \emph {et~al.}(2013)\citenamefont
  {Kshirsagar}, \citenamefont {Shaik}, \citenamefont {Krishna},\ and\
  \citenamefont {Tewari}}]{kshirsagar_photoinduced_2013}%
  \BibitemOpen
  \bibfield  {author} {\bibinfo {author} {\bibfnamefont {S.~D.}\ \bibnamefont
  {Kshirsagar}}, \bibinfo {author} {\bibfnamefont {U.~P.}\ \bibnamefont
  {Shaik}}, \bibinfo {author} {\bibfnamefont {M.~G.}\ \bibnamefont {Krishna}},\
  and\ \bibinfo {author} {\bibfnamefont {S.~P.}\ \bibnamefont {Tewari}},\
  }\href {https://doi.org/10.1007/s00339-012-7300-0} {\bibfield  {journal}
  {\bibinfo  {journal} {Applied Physics A}\ }\textbf {\bibinfo {volume}
  {111}},\ \bibinfo {pages} {861} (\bibinfo {year} {2013})}\BibitemShut
  {NoStop}%
\bibitem [{\citenamefont {Vasileiadis}\ and\ \citenamefont
  {Yannopoulos}(2014)}]{vasileiadis_photo-induced_2014}%
  \BibitemOpen
  \bibfield  {author} {\bibinfo {author} {\bibfnamefont {T.}~\bibnamefont
  {Vasileiadis}}\ and\ \bibinfo {author} {\bibfnamefont {S.~N.}\ \bibnamefont
  {Yannopoulos}},\ }\href {https://doi.org/10.1063/1.4894868} {\bibfield
  {journal} {\bibinfo  {journal} {Journal of Applied Physics}\ }\textbf
  {\bibinfo {volume} {116}},\ \bibinfo {pages} {103510} (\bibinfo {year}
  {2014})}\BibitemShut {NoStop}%
\bibitem [{\citenamefont {Woodbury}\ \emph {et~al.}(2020)\citenamefont
  {Woodbury}, \citenamefont {Schwartz}, \citenamefont {Rockafellow},
  \citenamefont {Wahlstrand},\ and\ \citenamefont {Milchberg}}]{PRL}%
  \BibitemOpen
  \bibfield  {author} {\bibinfo {author} {\bibfnamefont {D.}~\bibnamefont
  {Woodbury}}, \bibinfo {author} {\bibfnamefont {R.~M.}\ \bibnamefont
  {Schwartz}}, \bibinfo {author} {\bibfnamefont {E.}~\bibnamefont
  {Rockafellow}}, \bibinfo {author} {\bibfnamefont {J.~K.}\ \bibnamefont
  {Wahlstrand}},\ and\ \bibinfo {author} {\bibfnamefont {H.~M.}\ \bibnamefont
  {Milchberg}},\ }\href {https://doi.org/10.1103/PhysRevLett.124.013201}
  {\bibfield  {journal} {\bibinfo  {journal} {Phys. Rev. Lett.}\ }\textbf
  {\bibinfo {volume} {124}},\ \bibinfo {pages} {013201} (\bibinfo {year}
  {2020})}\BibitemShut {NoStop}%
\bibitem [{\citenamefont {Oishi}\ \emph {et~al.}(1987)\citenamefont {Oishi},
  \citenamefont {Okamoto},\ and\ \citenamefont {Sunada}}]{OISHI198729}%
  \BibitemOpen
  \bibfield  {author} {\bibinfo {author} {\bibfnamefont {K.}~\bibnamefont
  {Oishi}}, \bibinfo {author} {\bibfnamefont {K.}~\bibnamefont {Okamoto}},\
  and\ \bibinfo {author} {\bibfnamefont {J.}~\bibnamefont {Sunada}},\ }\href
  {https://doi.org/https://doi.org/10.1016/0040-6090(87)90118-0} {\bibfield
  {journal} {\bibinfo  {journal} {Thin Solid Films}\ }\textbf {\bibinfo
  {volume} {148}},\ \bibinfo {pages} {29 } (\bibinfo {year}
  {1987})}\BibitemShut {NoStop}%
\bibitem [{\citenamefont {Bonse}\ \emph {et~al.}(2009)\citenamefont {Bonse},
  \citenamefont {Rosenfeld},\ and\ \citenamefont
  {Kr{\"u}ger}}]{bonse_role_2009}%
  \BibitemOpen
  \bibfield  {author} {\bibinfo {author} {\bibfnamefont {J.}~\bibnamefont
  {Bonse}}, \bibinfo {author} {\bibfnamefont {A.}~\bibnamefont {Rosenfeld}},\
  and\ \bibinfo {author} {\bibfnamefont {J.}~\bibnamefont {Kr{\"u}ger}},\
  }\href {https://doi.org/10.1063/1.3261734} {\bibfield  {journal} {\bibinfo
  {journal} {Journal of Applied Physics}\ }\textbf {\bibinfo {volume} {106}},\
  \bibinfo {pages} {104910} (\bibinfo {year} {2009})}\BibitemShut {NoStop}%
\bibitem [{\citenamefont {Huang}\ and\ \citenamefont {Xu}(2014)}]{huang2014}%
  \BibitemOpen
  \bibfield  {author} {\bibinfo {author} {\bibfnamefont {M.}~\bibnamefont
  {Huang}}\ and\ \bibinfo {author} {\bibfnamefont {Z.}~\bibnamefont {Xu}},\
  }\href {https://doi.org/10.1002/lpor.201300212} {\bibfield  {journal}
  {\bibinfo  {journal} {Laser \& Photonics Reviews}\ }\textbf {\bibinfo
  {volume} {8}},\ \bibinfo {pages} {633} (\bibinfo {year} {2014})}\BibitemShut
  {NoStop}%
\bibitem [{\citenamefont {Sokolowski‐Tinten}\ \emph
  {et~al.}(2010)\citenamefont {Sokolowski‐Tinten}, \citenamefont {Barty},
  \citenamefont {Boutet}, \citenamefont {Shymanovich}, \citenamefont {Chapman},
  \citenamefont {Bogan}, \citenamefont {Marchesini}, \citenamefont
  {Hau‐Riege}, \citenamefont {Stojanovic}, \citenamefont {Bonse},
  \citenamefont {Rosandi}, \citenamefont {Urbassek}, \citenamefont {Tobey},
  \citenamefont {Ehrke}, \citenamefont {Cavalleri}, \citenamefont
  {D{\"u}sterer}, \citenamefont {Redlin}, \citenamefont {Frank}, \citenamefont
  {Bajt}, \citenamefont {Schulz}, \citenamefont {Seibert}, \citenamefont
  {Hajdu}, \citenamefont {Treusch}, \citenamefont {Bostedt}, \citenamefont
  {Hoener},\ and\ \citenamefont {M{\"o}ller}}]{sokolowski_2010}%
  \BibitemOpen
  \bibfield  {author} {\bibinfo {author} {\bibfnamefont {K.}~\bibnamefont
  {Sokolowski‐Tinten}}, \bibinfo {author} {\bibfnamefont {A.}~\bibnamefont
  {Barty}}, \bibinfo {author} {\bibfnamefont {S.}~\bibnamefont {Boutet}},
  \bibinfo {author} {\bibfnamefont {U.}~\bibnamefont {Shymanovich}}, \bibinfo
  {author} {\bibfnamefont {H.}~\bibnamefont {Chapman}}, \bibinfo {author}
  {\bibfnamefont {M.}~\bibnamefont {Bogan}}, \bibinfo {author} {\bibfnamefont
  {S.}~\bibnamefont {Marchesini}}, \bibinfo {author} {\bibfnamefont
  {S.}~\bibnamefont {Hau‐Riege}}, \bibinfo {author} {\bibfnamefont
  {N.}~\bibnamefont {Stojanovic}}, \bibinfo {author} {\bibfnamefont
  {J.}~\bibnamefont {Bonse}}, \bibinfo {author} {\bibfnamefont
  {Y.}~\bibnamefont {Rosandi}}, \bibinfo {author} {\bibfnamefont {H.~M.}\
  \bibnamefont {Urbassek}}, \bibinfo {author} {\bibfnamefont {R.}~\bibnamefont
  {Tobey}}, \bibinfo {author} {\bibfnamefont {H.}~\bibnamefont {Ehrke}},
  \bibinfo {author} {\bibfnamefont {A.}~\bibnamefont {Cavalleri}}, \bibinfo
  {author} {\bibfnamefont {S.}~\bibnamefont {D{\"u}sterer}}, \bibinfo {author}
  {\bibfnamefont {H.}~\bibnamefont {Redlin}}, \bibinfo {author} {\bibfnamefont
  {M.}~\bibnamefont {Frank}}, \bibinfo {author} {\bibfnamefont
  {S.}~\bibnamefont {Bajt}}, \bibinfo {author} {\bibfnamefont {J.}~\bibnamefont
  {Schulz}}, \bibinfo {author} {\bibfnamefont {M.}~\bibnamefont {Seibert}},
  \bibinfo {author} {\bibfnamefont {J.}~\bibnamefont {Hajdu}}, \bibinfo
  {author} {\bibfnamefont {R.}~\bibnamefont {Treusch}}, \bibinfo {author}
  {\bibfnamefont {C.}~\bibnamefont {Bostedt}}, \bibinfo {author} {\bibfnamefont
  {M.}~\bibnamefont {Hoener}},\ and\ \bibinfo {author} {\bibfnamefont
  {T.}~\bibnamefont {M{\"o}ller}},\ }\href {https://doi.org/10.1063/1.3507123}
  {\bibfield  {journal} {\bibinfo  {journal} {AIP Conference Proceedings}\
  }\textbf {\bibinfo {volume} {1278}},\ \bibinfo {pages} {373} (\bibinfo {year}
  {2010})}\BibitemShut {NoStop}%
\bibitem [{\citenamefont {Marini}\ \emph {et~al.}(2012)\citenamefont {Marini},
  \citenamefont {Chermisi}, \citenamefont {Lavagnini}, \citenamefont
  {Di~Castro}, \citenamefont {Petrillo}, \citenamefont {Degiorgi},
  \citenamefont {Scandolo},\ and\ \citenamefont
  {Postorino}}]{marini_high-pressure_2012}%
  \BibitemOpen
  \bibfield  {author} {\bibinfo {author} {\bibfnamefont {C.}~\bibnamefont
  {Marini}}, \bibinfo {author} {\bibfnamefont {D.}~\bibnamefont {Chermisi}},
  \bibinfo {author} {\bibfnamefont {M.}~\bibnamefont {Lavagnini}}, \bibinfo
  {author} {\bibfnamefont {D.}~\bibnamefont {Di~Castro}}, \bibinfo {author}
  {\bibfnamefont {C.}~\bibnamefont {Petrillo}}, \bibinfo {author}
  {\bibfnamefont {L.}~\bibnamefont {Degiorgi}}, \bibinfo {author}
  {\bibfnamefont {S.}~\bibnamefont {Scandolo}},\ and\ \bibinfo {author}
  {\bibfnamefont {P.}~\bibnamefont {Postorino}},\ }\href
  {https://doi.org/10.1103/PhysRevB.86.064103} {\bibfield  {journal} {\bibinfo
  {journal} {Physical Review B}\ }\textbf {\bibinfo {volume} {86}},\ \bibinfo
  {pages} {064103} (\bibinfo {year} {2012})}\BibitemShut {NoStop}%
\bibitem [{\citenamefont {Mellor}\ and\ \citenamefont
  {Bain}(2006)}]{mellor_array_2006}%
  \BibitemOpen
  \bibfield  {author} {\bibinfo {author} {\bibfnamefont {C.~D.}\ \bibnamefont
  {Mellor}}\ and\ \bibinfo {author} {\bibfnamefont {C.~D.}\ \bibnamefont
  {Bain}},\ }\href {https://doi.org/10.1002/cphc.200500348} {\bibfield
  {journal} {\bibinfo  {journal} {ChemPhysChem}\ }\textbf {\bibinfo {volume}
  {7}},\ \bibinfo {pages} {329} (\bibinfo {year} {2006})}\BibitemShut {NoStop}%
\bibitem [{\citenamefont {Bonch-Bruevich}\ \emph {et~al.}(1992)\citenamefont
  {Bonch-Bruevich}, \citenamefont {Libenson}, \citenamefont {Makin},\ and\
  \citenamefont {Trubaev}}]{bruevich}%
  \BibitemOpen
  \bibfield  {author} {\bibinfo {author} {\bibfnamefont {A.~M.}\ \bibnamefont
  {Bonch-Bruevich}}, \bibinfo {author} {\bibfnamefont {M.~N.}\ \bibnamefont
  {Libenson}}, \bibinfo {author} {\bibfnamefont {V.~S.}\ \bibnamefont
  {Makin}},\ and\ \bibinfo {author} {\bibfnamefont {V.~V.}\ \bibnamefont
  {Trubaev}},\ }\href {https://doi.org/10.1117/12.56133} {\bibfield  {journal}
  {\bibinfo  {journal} {Optical Engineering}\ }\textbf {\bibinfo {volume}
  {31}},\ \bibinfo {pages} {718 } (\bibinfo {year} {1992})}\BibitemShut
  {NoStop}%
\bibitem [{\citenamefont {Liao}\ \emph {et~al.}(2015)\citenamefont {Liao},
  \citenamefont {Ni}, \citenamefont {Qiao}, \citenamefont {Huang},
  \citenamefont {Bellouard}, \citenamefont {Sugioka},\ and\ \citenamefont
  {Cheng}}]{Liao2015}%
  \BibitemOpen
  \bibfield  {author} {\bibinfo {author} {\bibfnamefont {Y.}~\bibnamefont
  {Liao}}, \bibinfo {author} {\bibfnamefont {J.}~\bibnamefont {Ni}}, \bibinfo
  {author} {\bibfnamefont {L.}~\bibnamefont {Qiao}}, \bibinfo {author}
  {\bibfnamefont {M.}~\bibnamefont {Huang}}, \bibinfo {author} {\bibfnamefont
  {Y.}~\bibnamefont {Bellouard}}, \bibinfo {author} {\bibfnamefont
  {K.}~\bibnamefont {Sugioka}},\ and\ \bibinfo {author} {\bibfnamefont
  {Y.}~\bibnamefont {Cheng}},\ }\href {https://doi.org/10.1364/OPTICA.2.000329}
  {\bibfield  {journal} {\bibinfo  {journal} {Optica}\ }\textbf {\bibinfo
  {volume} {2}},\ \bibinfo {pages} {329} (\bibinfo {year} {2015})}\BibitemShut
  {NoStop}%
\bibitem [{\citenamefont {Yang}\ \emph {et~al.}(2009)\citenamefont {Yang},
  \citenamefont {Liu},\ and\ \citenamefont {Lalanne}}]{Lalanne2009}%
  \BibitemOpen
  \bibfield  {author} {\bibinfo {author} {\bibfnamefont {X.~Y.}\ \bibnamefont
  {Yang}}, \bibinfo {author} {\bibfnamefont {H.~T.}\ \bibnamefont {Liu}},\ and\
  \bibinfo {author} {\bibfnamefont {P.}~\bibnamefont {Lalanne}},\ }\href
  {https://doi.org/10.1103/PhysRevLett.102.153903} {\bibfield  {journal}
  {\bibinfo  {journal} {Phys. Rev. Lett.}\ }\textbf {\bibinfo {volume} {102}},\
  \bibinfo {pages} {153903} (\bibinfo {year} {2009})}\BibitemShut {NoStop}%
\end{thebibliography}%

\end{document}